\documentclass[nonblindrev,twocolumn]{informs3} 

\DoubleSpacedXI 

\usepackage{endnotes}
\let\footnote=\endnote

\usepackage{natbib}
 \bibpunct[, ]{(}{)}{,}{a}{}{,}%
 \def\bibfont{\small}%
 \def\newblock{\ }%
 \usepackage{rotating}

 \def\bibfont{\small}%
 \def\newblock{\ }%
\newcolumntype{d}[1]{D{.}{.}{#1}}

\newcommand{\I}{{\cal I}}

\def\R{I\!\!R}

\TheoremsNumberedThrough     
\ECRepeatTheorems

\EquationsNumberedThrough    

\usepackage{amsmath}

\usepackage[linesnumbered,ruled,vlined]{algorithm2e}

\begin{document}


\RUNAUTHOR{J.F. Monge}
\RUNTITLE{Cardinality constrained portfolio selection  via  factor models}

\TITLE{Cardinality constrained portfolio selection via  factor models}

\ARTICLEAUTHORS{%
\AUTHOR{Juan Francisco Monge}
\AFF{Centro de Investigaci\'{o}n Operativa, Universidad Miguel Hern\'{a}ndez de  Elche, (Spain), \EMAIL{monge@umh.es}} 
} 

\ABSTRACT{%

In this paper we propose and discuss different 0-1 linear models in order to solve the cardinality constrained portfolio problem  by using   factor models.    Factor models are used to build  portfolios to track indexes, together with other objectives, also need a smaller number   of parameters to estimate than the classical Markowitz model. The addition of the cardinality constraints limits the number of securities in the portfolio. Restricting the number of securities in the portfolio allows us to obtain a  concentrated  portfolio, reduce the risk and limit   transaction costs. To solve this  problem, a pure 0-1 model is presented in this work, the 0-1 model is constructed by means of  a piecewise linear approximation.   
We also present a new quadratic combinatorial problem, called a minimum edge-weighted clique problem, to  obtain an equality weighted cardinality constrained portfolio. A piecewise linear approximation for this problem is presented in the context of  a multi factor model.  For a single factor model, we present a fast heuristic, based on   some theoretical results  to obtain an equality weighted cardinality constraint portfolio. 
The consideration of a piecewise linear approximation allow us to reduce significantly   the computation time required for the equivalent quadratic problem.   
 Computational results from the  0-1 models are compared to those  using a  state-of-the-art Quadratic MIP solver. 

}
\KEYWORDS{finance, portfolio selection, Factor models, minimum-variance portfolio.}

\maketitle

\section{Introduction}

The portfolio selection problem deals with selecting   a collection of financial assets and in what proportion, according to the investor's risk preference, with the aim of  obtaining the  maximum expected return. 

The selection of assets  allocated to the portfolio can be managed using different approaches:   minimum risk allocation, equal weighting, risk parity, Sharpe ratio, and many others. 

In the seminar work of Markowitz (\citeyear{markowitz52}), the return and risk  are evaluated by means of the expected value and the variance of the  selected assets. Markowitz introduced the concept of an efficient frontier and showed that there is a set of optimal portfolios, not only one. The classical Markowitz model can be formulated as a quadratic linear model, and the investor can find an optimal portfolio maximizing the expected return under a  risk level, $  w^* = \arg_{w} \max \{ \,  w^T \mu  \,\, s.t. \,\, w^T \Sigma w= \sigma^* ,\, \,  w^T1=1\}$, or minimizing  the risk  under a return level,  $  w^* = \arg_{w} \min \{ \,  w^T \Sigma w  \,\, s.t. \,\, w^T \mu =r^* ,\,\,  w^T1=1\}$, where $w$ denotes the vector of weights in the portfolio, $\mu$ the vector of expected returns, and $\Sigma$ the covariance matrix of expected returns.  A significantly important portfolio is given when the constraint related to the return level is relaxed, obtaining the global minimum risk  solution. This solution is important in the literature. For example, in \citep{demiguel2009_2} the authors show that the minimum variance portfolio is a  more reliable and robust outsample than the traditional mean variance portfolios.  Another important portfolio is given when a  tradeoff objective function return/risk is considered, $w^* = \arg_{w} \max \{ \, w^T \Sigma w  \, - \lambda \,  w^T \Sigma w  \,\, s.t. \,\,   w^T1=1\}$, where $\lambda$ is the risk  aversion coefficient. Although we have considered in this paper the minimum variance portfolio, we will see that the results can easily be applied  to the objective functions mentioned above.

The factor model theory establishes the expected return of each asset as a linear function on   
the risk factors, through the parameter $\beta$, where $\beta$ is a measure of the risk contribution for the individual asset to the portfolio. The father of factor models is W.F. Sharpe, and their Capital Asset Pricing Model (CAPM) theory, see \citep{sharpe63,sharpe64}. 

The Markowitz mean-variance framework requires that are  estimate a large number of parameters. If there are $n$ assets, we need to estimate  $n$ means, $n$ variances and $n(n-1)/2$ covariances, $0(n^2)$. The factor model requires fewer parameters to be estimated;  the order is given by the number of factors $m$, i.e. $O(m^2)$, where the number of factors  $m$ is much smaller than $n$.

The cardinality constrained portfolio problem is a classic problem in the literature.  In \citep{chang2000} the authors present several properties for the efficient frontier for the cardinality constrained problem in the classical mean-variance Markowitz model, giving properties of solutions,  showing  for example, the discontinuity of the efficient frontier,  also  as the traditional minimization of trade-of objective function mean/risk does not provide all the efficient solutions. The  authors  also present different heuristics for this problem, while  \citep{Beasley2011} is related to  methaheuristic approaches. The exact resolution of the problem is analyzed in \citep{cesarone2013}, where the authors present an exact algorithm for medium size problems, that provide a good approximation for larger problems.

In \citep{Shaw2008} the authors present a Lagrangian decomposition scheme for the cardinality constrained portfolio problem. The authors present a decomposition of the covariance matrix in two matrix; a diagonal matrix with the risk of each asset, and another  non-diagonal  with the covariance among the factors. This  idea allows them to reduce the dimensions of the quadratic problem to be solved. See  \citep{gao2013} for another application of the  Lagrangian decomposition scheme for this  problem. See in   \citep{Bertsimas} an alternative procedure based on solving a succession of problems into  a tree search.

Another  alternative that can be found in the literature, regarding the cardinality constrained problem,  refers to the investment being  made in lots,   the excess  capital  going to a risk-free asset, see \citep{duanli2006}.  In \citep{justo2011} the authors propose a algebraic  algorithm to solve the integer problem with linear  objective function,  the expected return,  under linear and  non-linear constraints.

All the above papers  only deal with the classical Markowitz model; these papers do not integrate the cardinality constrain in factor models. To the best  of our knowledge there does not exist in the literature a paper combining factor models  and the cardinality constraint.

The main contribution of this work relates to   the  linear approximation of the quadratic factor model problem. Two  linear approximations are considered in this work;  the first  through a piecewise linear function, and the second  imposing the  equal weighted in the solution. The singularity  present in the  covariance matrix of the  factor models allows us to take advantage  above the Cplex solver.

The rest of paper is organized as follows. Section 2 deals with the main concepts of factor models and introduces the mathematical notation for the cardinality constrained minimum variance problem via factor models, the piecewise linear approximation of this problem and the model where  the  equal weighted constraint is imposed. Section 3 studies the problem  where a single factor is considered; it also presents theoretical results for this new combinatorial problem and a heuristic algorithm to solve it. Section 4 reports the computational results for a set of instances used in the literature. Finally, section 5 concludes and outlines future plans.  

\section{Factor Model}

For a risky  asset $i\in I$, a factor model assumes that the return rates $r_i$ of asset $i$ is given by 
 $r_i = \alpha_i + \beta_i F +\epsilon_i$, 
where $F=(f_1,\dots,f_m)$ is a vector of random variables called factors, with $E(f_l )=0$, $\alpha_i\in \R$  is a constant, $\beta_i\in \R^m$ is a constant vector and $\epsilon_i$ is a (error) mean zero random variable, uncorrelated with the factors, $E(\epsilon_i)=0$ and $E(\epsilon_i \cdot  f_l)=0$. The factors $F$ are correlated with covariance matrix $\Sigma_F$. We use the notation $\sigma_{lm}=E(f_l\cdot f_m)$ and $\sigma_{\epsilon_i}^2=E(\epsilon_i^2)$. 

For a portfolio formed with $n$ assets, defined by weights $w^T=$($w_1$, $w_2$, \dots, $w_n$), then the portfolio is determined by a factor model, where the return $r=\sum_{i\in I} w_i r_i$ of the porfolio is $$ r= \sum_{i\in I} w_i \alpha_i + \sum_{i\in I} w_i  \beta_i^T F +\sum_{i\in I} w_i \epsilon_i $$

In matrix form:
$$ r= w^T (\alpha +   \beta^T F + \epsilon )$$

where

$ \alpha \in \R^n$, $\beta \in \R^{m\times n}$, $F \sim N(0,\Sigma_F)$, $\epsilon \sim N(0,\Sigma_\epsilon)$ 

The mean-variance parameters can be calculated directly in terms of the factor model:

$$ E(r)= w^T \alpha=\sum_{i=1}^n w_i \alpha_i$$
$$V(r)=w^T \Sigma_r w = w^T (\beta^T \Sigma_F \beta +  \Sigma_\epsilon )w=  w^T \beta^T \Sigma_F \beta w +  w^T\Sigma_\epsilon w=$$
$$ = \sum_{i,j \in I}  \sum_{l,m \in F}  w_i w_j \beta_{il} \beta_{jm} \sigma_{lm} +  \sum_{i\in I} w_i ^2 \sigma_{\epsilon_i}^2 $$

\subsection{Cardinality constrained minimum-variance portfolio problem with  factor models.}

Let $K$ be the desired number of assets in the portfolio. Consider the following decision variables:
\begin{description}
\item[$x_i$,] binary variable that takes value 1 if the asset $i$ is selected, $\forall i\in I$. 
\item[$w_i$,] weight of asset $i$ in the portfolio, $\forall i\in I$. 
\end{description}

Then, the Cardinality Constrained Minimum Variance portfolio via    Factor Models (CCMVFM) is the solution to the mixed 0-1 binary quadratic optimization problem:

\begin{equation}\label{CCMVFM}
\begin{array}{r@{}ll}
(CCMVFM) \quad & \displaystyle \quad\min_{w,x} \quad \qquad & \displaystyle  \sum_{i,j \in I}  \sum_{l,m \in F}   \beta_{il} \beta_{jm} \sigma_{rl} w_i w_j+  \sum_{i\in I}  \sigma_{\epsilon_i}^2 w_i ^2\\ 
& \quad s.t.   &\displaystyle \sum_{i\in I} w_i =1\, ,  \qquad \\ 
		&&\displaystyle  \sum_{i\in I}  x_i \leq K,  \\
		&&\displaystyle 0 \leq w_i \leq x_i\, ,  \qquad  \forall i\in I ,  \\
		&&\displaystyle x_i \in \{0,1\}\, , \qquad  \forall i\in I .
\end{array}
\end{equation}

If the factors are uncorrelated ($\sigma_{lm}=0, \forall \, l,m \in F : l\neq m$), the objective function of the problem ($CCMVFM$ (\ref{CCMVFM})) can be written as:

$$ \displaystyle \quad\min_{w,x} \quad   \displaystyle  \sum_{i,j \in I}  \sum_{l \in F} \beta_{il} \beta_{jl} \sigma_{ll}  w_i w_j +  \sum_{i\in I}  \sigma_{\epsilon_i}^2 w_i ^2$$

\subsection*{Piecewise linear approximation}

In order to improve the computational time required to solve the $CCMVFM$ model (\ref{CCMVFM}), we propose a piecewise linear approximation. Consider $S_w$, set of $s$ ordered disjoint segments of variable $w_i$, i.e, the set of ordered disjoints segments in the interval $[0,1]=[\overline{w}_i^0=0,\overline{w}_i^1)\cup[\overline{w}_i^1,\overline{w}_i^2)\cup \cdots \cup [\overline{w}_i^{s-1},\overline{w}_i^s=1]$; and,  $S_\beta$, set of $t$ ordered disjoint segments of variable $\beta_{\cdot l}$ in the interval $[\beta_{\min},\beta_{\max}]=[\overline{\beta}_{\cdot l}^0=\beta_{\min},\overline{\beta}_{\cdot l}^1)\cup[\overline{\beta}_{\cdot l}^1,\overline{\beta}_{\cdot l}^2)\cup \cdots \cup [\overline{\beta}_{\cdot l}^{t-1},\overline{\beta}_{\cdot l}^t=\beta_{\max}]$, where $\beta_{\cdot l}=\sum_{i\in I} w_i \beta_{il}$. So, the  quadratic model (\ref{CCMVFM}) can be  approximated by the following 0-1 pure quadratic  model:

\begin{align}
 (CCMVFM_{LA})&&  \displaystyle  \min_{x,y}  \qquad  &\displaystyle  \sum_{t, t' \in S_{\beta}} \sum_{l,m \in F} \sigma_{lm} \overline{\beta}_{\cdot l}^t \overline{\beta}_{\cdot m}^{t'}  y_l^ty_m^{t'} + \sum_{i\in I, s\in S_w} \, \sigma_{\epsilon_i}^2 \, {\overline{w}_{i}^s}^2 \,   x_{i}^s  \\ 
  &&s.t. \qquad &\displaystyle \sum_{s\in S_w} x_i^s =1\, ,  \qquad \forall i\in I  , \\
    & &       & \displaystyle  \sum_{t\in S_\beta}  \overline{y}^t_{ l}=1 \, , \qquad \forall l\in F,  \\
     &&       &\displaystyle \sum_{i\in I}\sum_{s\in S_w: s>0} x_i^s \leq K\, ,  \\
      &&       &\displaystyle\sum_{i\in I}\sum_{s\in S_w: s>0} \overline{w}_{i}^s x_i^s =1\, ,  \\
       &&     &\displaystyle \sum_{i\in I}\sum_{s\in S_w: s>0} \beta_{il} \overline{w}_{i}^s x_i^s  \leq   \sum_{t\in S_\beta}{\overline{\beta}^t_{\cdot l} y_l^t}\, , \qquad \forall l\in F , \\
       &&     &\displaystyle x_i^s  \in\{0,1\},\qquad  \forall i\in I \, ,  s\in S_w, \\ 
       &&     & \displaystyle  y_l^{t}  \in\{0,1\}\, , \qquad  \forall l\in F, t \in S_\beta , 
\end{align}

where the 0-1 variable  $x_i^s$ takes value 1 if the weight of asset $i$ is fixed in the solution at level $\overline{w}_i^s$, and the 0-1 variable   $ y_l^{t} $ takes value 1 if $\overline{\beta}_{\cdot l}^{t}$ is the least  upper bound of  $\beta_{\cdot l}$ in the set $S_\beta$. 

If  the factors are uncorrelated ($\sigma_{lm}=0, \forall \, l,m\in F: l\neq m$), then the quadratic model (2)-(9) becomes in the following linear pure  0-1 model: 

\begin{align}
  \displaystyle  \min_{x,y}  \qquad&   \sum_{t \in S_{\beta}} \sum_{l \in F} \sigma_{ll} {\overline{\beta}_{\cdot l}^t}^2 y_l^t  +  \displaystyle \sum_{i\in I, s\in S_w} \, \sigma_{\epsilon_i}^2 \, {\overline{w}_{i}^s}^2 \,   x_{i}^s  \\
  s.t. \qquad &(3)-(9). \nonumber
\end{align}

\subsection{Equality weighted cardinality constrained portfolio problem}

A simplification model of ($CCMVFM$ (\ref{CCMVFM})) is the model when the equality weighted constraint is imposed, i.e., the weight of asset $i$, $w_i$, is $1/K$ if the asset $i$ is selected, and  0 otherwise. The problem of finding  find the best Equality Weighted Cardinality Constrained Minimum Variance portfolio for a multi Factor Model (EWCCMVFM), i.e., the solution of  CCMVFM  problem, when the weight of all assets selected are the same, is the solution of the 0-1 pure binary quadratic optimization problem:

 \begin{equation}\label{EWCCMVFM}
\begin{array}{r@{}ll}
(EWCCMVFM) \quad & \displaystyle \quad \frac{1}{K^2} \min_{x} \quad \qquad & \displaystyle  \sum_{i,j \in I}  \sum_{l,m \in F}   \beta_{il} \beta_{jm} \sigma_{lm} x_i x_j+  \sum_{i\in I}  \sigma_{\epsilon_i}^2 x_i ^2 \\ 
& \quad s.t.   &\displaystyle  \sum_{i\in I}  x_i = K \, , \\
		&&\displaystyle x_i \in \{0,1\}\, ,  \qquad  \forall i\in I,
\end{array}
\end{equation}
where $x_i$ takes value 1 if the asset $i$ is selected, and 0 otherwise. The constraint ($\sum_{i\in I} w_i =1$) in (\ref{CCMVFM}) forces us to select exactly  $K$ assets  ($\sum_{i \in I}x_i=K$)  in the model (\ref{EWCCMVFM}), i.e., we need to impose the equality in the cardinality constraint. Note also, we can replace the term $\sum_{i\in I}  \sigma_{\epsilon_i}^2 x_i ^2$ in the objective function for $\sum_{i\in I}  \sigma_{\epsilon_i}^2 x_i $, because $x_i$ takes the value 0 or 1.

The problem  (\ref{EWCCMVFM}) can be written as $\{  \min_x \, \sum_{i,j \in I} a_{ij} x_ix_j,\,\, \text{s.t.}   \sum_{i\in I}  x_i = K ,\,\, x_i \in\{0,1\}\, \forall i\in I \}$, where  

$a_{ij}=\left\{\begin{array}{ll} 
\frac{1}{K^2}\sum_{l,m \in F} \beta_{il} \beta_{jm} \sigma_{lm} & \quad \text{if} \quad  i\neq j, \\
\frac{1}{K^2}\sum_{l,m \in F} \beta_{il} \beta_{im} \sigma_{lm} +  \sigma_{\epsilon_i}^2  & \quad \text{if} \quad  i=j.
\end{array}\right .$\\

A well-know problem in the literature is the Maximum Edge-Weigted Clique Problem (MEWCP), see \citep{glover2007,souza2000} among others. The MEWCP problem can be defined as follows: Given a complete graph $G=(V,E)$ with nodes and unrestricted edge weights $c_{ij}$, find a subclique of $G$ with $k$  nodes such that the sum of the weights in the sub-clique is maximized. A non-linear formulation of the problem is:

\begin{equation}\label{MEWCP}
\begin{array}{r@{}ll}
(MEWCP) \quad &\quad\max  \qquad & \displaystyle  \sum_{i,j \in V, i<j} c_{ij} x_{i}x_{j} \\ 
 &\quad s.t.   \qquad&\displaystyle \sum_{i\in V} x_{i} \leq k,   \\ 
 & &  \displaystyle  x_i \in \{0,1\},  \qquad  \forall i \in I.
\end{array}
\end{equation}

\begin{proposition} An instance of  EWCCMVFM  problem can be transformed into  an instance of  MEWCP.
\end{proposition}
\proof{Proof.} 
Let $\cal G$ a larga number, for example $\cal G=\max\{a_{ij}, \,\, i,j\in I\}$, then, the solution of the problem ($MEWCP$) with  
$c_{ij}=\left\{\begin{array}{ll} 
\cal G -(2\, a_{ij}+\displaystyle \frac{a_{ii}+a_{jj}}{K-1})& \quad \text{if} \quad  i< j \\
0  & \quad \text{if} \quad  i\geq j 
\end{array}\right .$\\
is solution of the problem  ($EWCCMVFM$).
\Halmos
\endproof

Proposition 1 implies that the EWCCMVFM problem inherits all the properties of  MEWCP. Nevertheless, the EWCCMVFM has  remarkable matrix coefficients, see appendix. This fact makes this problem (EWCCMVFM) more treatable computationally.

There exists in the literature linear formulations for the MEWCP, however these formulations are not considered in this work because they  behaved worse than the quadratic formulation (\ref{MEWCP}), see \citep{souza2000} and the references therein for a good explanation of the MEWCP problem.

\subsection*{Piecewise linear approximation}

Using the same approximation used in (CCMVFM), the problem (EWCCMVFM) can be approximated by  the following quadratic 0-1 problem:

\begin{align}
 (EWCCMVFM_{LA})&&  \displaystyle \frac{1}{K^2}  \min \quad  &  \sum_{t, t' \in S_{\beta}} \sum_{l,m \in F} \sigma_{lm} \overline{\beta}_{\cdot l}^t \overline{\beta}_{\cdot m}^{t'}  y_l^ty_m^{t'}  +\sum_{i\in I} \, \sigma_{\epsilon_i}^2 \,   x_{i} \\ 
  && s.t \qquad           &\sum_{i} x_i = K , \\
    &&        &\frac{1}{K}\sum_{i} \beta_{il}  x_i  \leq   \sum_t{\overline{\beta}^s_{\cdot l} y_l^t}\, , \qquad \forall l\in F , \\
     &&       &  \sum_{t\in S_\beta}  y^t_{ l}=1,  \qquad \forall l\in F , \\
    &&        &x_i  \in\{0,1\}, \qquad  \forall i\in I, \\
    &&        & y_l^{t}  \in\{0,1\} ,\qquad  \forall  l\in F, t \in S_\beta . 
\end{align}

If  the factors are uncorrelated ($\sigma_{lm}=0, \forall \, l,m\in F: l\neq m$), then the quadratic model (13)-(18) becomes in the following linear pure 0-1 model:

\begin{align}
 \quad \displaystyle \frac{1}{K^2}  \min \qquad  &\sum_{i\in I} \, \sigma_{\epsilon_i}^2 \,   x_{i}  + \sum_{t \in S_{\beta}} \sum_{l \in F} \sigma_{ll} {\overline{\beta}_{\cdot l}^t  }^2 y_l^t\\ 
   s.t \qquad           & (14)-(18) \nonumber
\end{align}

We have presented different models for the cardinality constrained portfolio selection via factor models: the $CCMVFM $ problem and its linear approximation ($CCMVFM_{LA}$), and the  $EWCCMVFM$ problem and its linear approximation ($EWCCMVFM_{LA}$). All these models have different classifications in mathematical programming theory depending on their characteristics: linear or non-linear objective function, continuous or integer variables, etc. Table  \ref{t:classification} shows  the characteristics of problems defined above, depending on whether the model considers correlated  or uncorrelated factors. Note that on  consideration of uncorrelated factors, both approximations  become in a 0-1 pure linear problems.

The  linear approximation of  $CCMVFM$ and  $CCMVFM_{LA}$ models, needs to add to the model  new binary variables, one binary variable $x_i^s$ for each asset $i\in I$ and each segment $s\in S_w$ considered,  and one variable $y_l^t$ for each factor $l\in F$ and each segment $t\in S_{\beta}$.  Table \ref{t:dimension}. shows the dimension of each model, where a column under heading $n01$ gives the number of binary variables of each model, the following column $nc$ gives the number of continuous variables, and finally the column $m$ gives the number of constraints.  These dimensions are given by $N$ the number of assets, $NF$ the number of factors, $ |S_w|$ the number of segments considered for  each variable $x_i^s$ and finally $ |S_{\beta}|$ the number of segments considered for each variable $y_l^t$ .  The number of segments considered in the computational experience has been fixed to 500, for $ |S_{\beta}|$, and as a function of the parameter of cardinality $K$, for $ |S_{w}|$.

Although the dimensions of linear approximations are much higher than the original quadratic model (CCMVFM), we will see that, given the great advance currently present in the optimization solvers for combinatorial problems, the resolution of these lineal models is much less expensive than the equivalent quadratic model.

\section{Equality weighted  cardinality constrained minimum variance portfolio problem for a single factor model.}

In this  section we study some properties for the EWCCMV problem where only one factor is considered. For a single factor $f$, the return rates $r_i$ of asset $i\in I$ is given by $r_i=\alpha_i+\beta_i f +\epsilon_i$, where  $E(f)=0$ and $E(f^2)=\sigma_f^2$.   

The quadratic 0-1 model for the Equality Weighted Cardinality Constraint Minimum Variance  portfolio with a Single Factor $f$  (EWCCMVSF) is:

\begin{equation}\label{EWCCMVSF}
\begin{array}{r@{}ll}
(EWCCMVSF) \quad & \displaystyle \quad \frac{1}{K^2} \,\min_{x} \quad \qquad & \displaystyle \sigma_{f}^2  \sum_{i,j \in I}     \beta_{i} \beta_{j}  x_i x_j +  \sum_{i\in I}   \sigma_{\epsilon_i}^2 x_i \\ 
& \quad s.t.   &\displaystyle  \sum_{i\in I}  x_i = K,  \\
		&&\displaystyle x_i \in \{0,1\} ,\qquad  \forall i\in I .
\end{array}
\end{equation}

Problem (\ref{EWCCMVSF}) can be written as: 

\begin{equation}\label{EWCCMVSF2}
\begin{array}{r@{}ll}
(EWCCMVSF) \quad 
 &  \displaystyle \quad \frac{1}{K^2} \, \min_{\overline{\beta}, \overline{\alpha}, x}  \quad& \displaystyle \overline{\beta}^2 +\overline{\alpha}  \\ 
& \quad s.t.   &\displaystyle  \sum_{i\in I}  x_i = K  ,\\
			&& \overline{\beta} =  \sigma_f \sum_{i \in I} ,    \beta_{i}  x_i, \\
			&&  \overline{\alpha} =  \sum_{i \in I},    \sigma_{\epsilon_i}^2  x_i , \\ 
		&&\displaystyle x_i \in \{0,1\} ,\qquad  \forall i\in I .
\end{array}
\end{equation}

\subsection{Theoretical results}

Let $A$ be the set of points  on the plane, $A=\left \{( \beta_{i} \sigma_f,\sigma_{\epsilon_i}^2), \, \forall  i\in I \right \}$, and the cardinality parameter $K$.

\begin{definition} The addition set of $A$, denoted by $A(K)$, is the set of all points generated by the addition of $K$ points from $A$. 
\[ A(K) =  \left \{ \sum_{a_i\in S \subset A} a_i, \, \forall S\subset A :  |S|=K  \right \} \]
\end{definition}

\begin{definition} Convex hull of set $A(K)$, denoted by $conv(A(K))$, is the set  of all convex combination of points generate by addition of    $K$ points in $A$,  that is: 
\[ conv(A(K)) =  \left \{ \sum_{i=1}^N x_i a_i \, : \, a_i\in A,\, x_i \in R,\, 0\leq x_i \leq 1 ,  \, \sum_{i=1}^N x_i =K  \right \}.\]
\end{definition}

The linear relaxation of problem (\ref{EWCCMVSF})  and  (\ref{EWCCMVSF2})  can be written as follows:

\begin{equation}\label{EWCCMVSF3}
\begin{array}{r@{}ll}
(EWCCMVSF) \quad 
 &  \displaystyle \quad \frac{1}{K^2} \, \min_{\overline{\beta}, \overline{\alpha}}  \quad& \displaystyle \overline{\beta}^2 +\overline{\alpha}  \\ 
& \quad s.t.   & (\overline{\beta}, \overline{\alpha})\in conv(A(K)). \\ 
\end{array}
\end{equation}

\begin{proposition} The optimal solution of (\ref{EWCCMVSF3}) is reached in the frontier of set  $conv(A(K))$.
\end{proposition}
\proof{Proof.} 
It remains  to show that  this proposition is true. 
\Halmos
\endproof

\begin{theorem}[Carath\'eodory, \citep{caratheodory1907}] For $S\subset \cal R^d$, if $x\in conv(S)$ then $x\in conv(T)$ for some $T\subset S,|T|\leq d+1$. 
\end{theorem}
\proof{Proof.} 
\Halmos
\endproof

The Carath\'eodory theorem establishes that any point in $conv(A(K))\subset \cal R^2$ can 
be represented as a convex combination of 3 points of $A(K)$. Note that the 3 points are from $A(K)$, and each point in $A(K)$ is the addition of $K$ points of $A$. The next corollary  restricts the Carath\'eodory theorem to the frontier of set  $conv(A(K))$.

\begin{corollary} The frontier of the polyhedron  $conv(A(K))\subset \cal R^2$ is formed for faces of dimension 0 and 1, then the solution of   (\ref{EWCCMVSF3}), ($\overline{\beta}^*$, $\overline{\alpha}^*$),   is a convex combination of two points  of $A(K)$. Assuming that there are no collinear points in the frontier of $conv(A(K))$.
\end{corollary}
\proof{Proof.} 
\Halmos
\endproof

From the corollary 1 it follows that the solution of (\ref{EWCCMVSF3}) is reached in one point of $A(K)$, or in the linear combination of two of them. One consequence of this result is that the solutions only have two or less fractional values. We establish this property in the following  proposition.

\begin{proposition} The solution of the problem  (\ref{EWCCMVSF})  contains at most two variables with a fractional value.  
\end{proposition}
\proof{Proof.} 
If the solution of (\ref{EWCCMVSF}) is reached in a vertex $v$ of $conv(A(K))$, this point is the addition of $K$ points of $A$, therefore, exist $S\subset A: |S|=K$ such that $v=\sum_{a_i\in S}a_i$, and $x_i=1$ if $i\in S$. 

If the solution is reached in a face of dimension 1, an arista of $conv(A(K))$, then the solution is a convex combination of two vertex, $v_1=\sum_{a_i\in S_1}a_i$ and $v_2=\sum_{a_i\in S_2}a_i$,  of $conv(A(K))$, the two vertex defining the arista. 

Suppose that $S_1\cup S_2>K+1$, i.e., $v_1$ and $v_2$ differ in two or more points from $A$. For example, 

$v_1 =a_1+a_2+a_5+\cdots+a_K+a_{K+1}+a_{K+2}$, and  $v_2=a_3+a_4+a_5+\cdots+a_K+a_{K+1}+a_{K+2}$.

The interior point $0.5 v_1 + 0.5 v_2 = 0.5(a_1+a_2)+0.5(a_3+a_4)+a_5+\cdots +a_K+a_{K+1}+a_{K+2}$ can  also be written 
$0.5(a_1+a_3)+0.5(a_4+a_4)+a_5+\cdots +a_K+a_{K+1}+a_{K+2}=0.5(a_1+a_3+a_5+\cdots+a_K+a_{K+1}+a_{K+2})+0.5(a_2+a_4+a_5+\cdots+a_K+a_{K+1}+a_{K+2})=0.5 z_1 + 0.5z_2$, where $z_1,z_2\in A(K)$. If $z_1$ and $z_2$ belong to the interior of  $A(K)$, then $0.5 v_1 + 0.5 v_2$ is an interior point, also a contradiction. 
If $z_1$ or $z_2$ are vertexs of $conv(A(K))$, then $v_1$, $v_2$ and $z_1$ (or $v_1$, $v_2$ and $z_2$) are collinear points, and this contradicts the supposition that there are no collinear  points in the frontier of $conv(A(K))$.

Therefore, a point in the frontier of $conv(A(K))$ is a linear combination at most two points of $A(K)$, and these two points of   $A(K)$  differ at most in one point from $A$. 
\Halmos
\endproof

{\bf Remark:} In the multi factor model the solution is also in the frontier, but in this case the dimension of polyhedral facets are less or equal to $|F|$, where $|F|$ is the number of factors.  In this case  the solution  is  a combination of $|F| +1 $ points (vertices) of $A(K)$, but now,  these points (vertices) do not have to be consecutive, consequently they can differ in more than one point from $A$. It will be seen in the computational experience that the resolution of the factor models problem requires a little  time, as in practice the solution of the  linear relaxation of EWCCMVFM problem has few fractional variables.

\subsection{Algorithm for the Equality weighted  cardinality constrained minimum variance portfolio problem for a single factor model}

As an alternative to the  EWCCMVSF model, in this section we introduce a new algorithm for obtaining a fast solution to  this model. The algorithm is based on the next proposition, proposition 4.

Given the set of assets $T$ of cardinality $K$, the objective function value in (\ref{EWCCMVSF}) (without the constant factor $1/K$) is:

$ obj(S)=   \sum_{i\in T}\sigma_{\epsilon_i}^2 +\sigma^2_f \sum_{i,j\in T} \beta_i \beta_j  = \sum_{i\in T}\sigma_{\epsilon_i}^2 +\sigma^2_f  \beta_T^2  $, where $\beta_T=\sum_{i\in T}\beta_i$.


Let  $S\cup\{i\}$ and  $S\cup\{j\}$ two sets of cardinality $K$,  differentiating  in  a single element, then: 
\begin{align}
 obj(S\cup\{i\}) - obj(S\cup\{j\}) =  \sigma_{\epsilon_i}^2 - \sigma_{\epsilon_j}^2 +\sigma^2_f (\beta_S + \beta_i)^2  -\sigma^2_f (\beta_S + \beta_j)^2,  \label{objS_i}
\end{align}
where $\beta_S=\sum_{i\in S}\beta_i.$ 

\begin{definition} We say that the asset $i$ is better than asset $j$ for   set $S$, $i<<_{S} j$, if  $obj(S\cup\{i\}) \leq obj(S\cup\{j\})$.
\end{definition}
\begin{proposition}  If exist   $S^*\subset I$ and  $S\subset I$ with  $i<<_{S^*} j$ and $(\beta_{S^*}-\beta_{S})(\beta_i-\beta_j)>0$, then  $i<<_{S} j$. \label{proposition3}
\end{proposition}
\proof{Proof.} 
If $i<<_{S^*} j$, then $ \sigma_{\epsilon_i}^2 - \sigma_{\epsilon_j}^2 +\sigma^2_f (\beta_{S^*} + \beta_i)^2  -\sigma^2_f (\beta_{S^*} + \beta_j)^2 <0 $.\\
Suppose for a contradiction  that $i<<_{S} j$ is not true, then $ \sigma_{\epsilon_j}^2 - \sigma_{\epsilon_i}^2 +\sigma^2_f (\beta_S + \beta_j)^2  -\sigma^2_f (\beta_S + \beta_i)^2 \leq 0 $.
By adding the above expressions, we obtain  $ \sigma^2_f (\beta_{S^*} + \beta_i)^2  -\sigma^2_f (\beta_{S^*} + \beta_j)^2  +\sigma^2_f (\beta_S + \beta_j)^2  -\sigma^2_f (\beta_S + \beta_i)^2 < 0 $, then  $(\beta_{S^*}-\beta_{S})(\beta_i-\beta_j)<0$, and we have a contradiction, and this proves that if $i<<_{S^*} j$ then $i<<_{S} j$, with $(\beta_{S^*}-\beta_{S})(\beta_i-\beta_j)>0$.
\Halmos
\endproof

 The previous proposition allows us to build a constructive heuristic  for the EWCCMVSF problem, see algorithm description  in Algorithm  1. 
 
 Let us describe the algorithm. 
  
 As the first step, the  algorithm starts with an initial solution, $S_0$, formed by the assets with less $\beta$-value. At the second step, the algorithm identify the asset $j^* \in S_0$, in the set of assets that are  selected in the current solution,   with the greatest contribution  in the objective function. Next,  identify the asset $i^*\in \I\setminus S_0$, in the assets that are not selected in the current solution, with the lower contribution.  So, if the testing is positive then an improvement of the solution value of model EWCCMVSF can be  performed locally by the algorithm from the current solution. Otherwise, the improvement to the current solution could not  be performed and the algorithm ends. 
 
%
 
 Although the algorithm does not guarantee finding the optimal solution to the problem (EWCCMVSF (\ref {EWCCMVSF})), let us justify its  good behaviour. It will also be seen later in the computational experience.
 
 The optimal solution of (\ref{EWCCMVSF}) is a set of $K$ assets, namely $S^*$. If the asset $i$ belong to $S^*$ then $i<<_{S^*\setminus\{i\}} j$, $\forall j\notin S^*$, i.e, the asset $i$ is better than any $j$, $j\notin S^*$, combined with the assets of $S^*\setminus\{i\}$. The algorithm starts with a set formed by the assets of lower $\beta$. For each asset $i$ present in the optima solution, $i\in S^*$ and not present  in $S_0$, it holds that ($\beta_{S^*} >\beta_{S_0}$) and ($\beta_i > \beta_j$) for all $j\in S_0\setminus S^*$.  Therefore , the asset $i$ improves the solution provided by $S_0$. It is easy to prove that the algorithm will find the optimal solution as long as it  removes  from the set $S_0$ an optimal asset $i\in S^*$.

In order to improve the solution provided for the algorithm 1, we have developed a second algorithm, see Algorithm \ref{improving}. It is possible that the parameter of cardinality imposed was large, obtaining a solution which is worse than for a smaller  number of assets. Algorithm 2 looks for the asset in the solution with the largest contribution,  and  it looks to see if by  removing the asset, an improvement  is obtained. The algorithm repeats the procedure while improving the solution.

\IncMargin{1.em}
\begin{algorithm}
\SetKwData{Left}{left}\SetKwData{This}{this}\SetKwData{Up}{up}
\SetKwFunction{Union}{Union}\SetKwFunction{FindCompress}{FindCompress}
\SetKwInOut{Input}{input}\SetKwInOut{Output}{output}
\Input{A set $I$ of $N$ ordered assets (less $\beta_i$ first, with $i\in I$), and a set $A=\left \{( \beta_{i} \sigma_f,\sigma_{\epsilon_i}^2), \, \forall  i\in I \right \}$.}
\Input{Parameter of cardinality $K$.}
\BlankLine
\emph{Let $S_0=\{1,2,\dots,k\}$ the set of the first $k$ assets of $I$.}\\
\Repeat{$j^* <<_{S_0\setminus \{j^*\}} i^*$}{
\For{$j\in S_0$}{
Calculate $obj(S_0)-obj(S\setminus\{j\})=\sigma_{\epsilon_j}^2+ \sigma^2_f\left (( \sum_{k\in S_0}\beta_k)^2 -(\sum_{k\in S_0:k\neq j}\beta_k)^2 \right)$
}\emph{}
 \emph{ Let $j^*=\arg_{j\in S_0} \max \{obj(S_0)-obj(S\setminus\{j\}) \}$, i.e., $j^*$ is the asset in $S_0\subset I$ with the greater contribution in the objective function.}\\
 \For{$i\in I\setminus S_0$}{
Calculate $obj(S_0)  -  obj(\{S_0\setminus\{j^*\}\}\cup \{i\})= \sigma_{\epsilon_j}^2 - \sigma_{\epsilon_i}^2 +\sigma^2_f (\beta_S + \beta_j)^2  -\sigma^2_f (\beta_S + \beta_i)^2$
}
 \emph{ Let $i^*=\arg_{i\in I\setminus S_0} \max \{obj(S_0)  -  obj(\{S_0\setminus\{j^*\}\}\cup \{i\}) \}$, i.e., $i^*$ is the asset in $I\setminus S_0$ with the lower contribution in the objective function when asset $j^*$ is removed from $S_0$.}\\
 
  \lIf{$i^* <<_{S_0\setminus \{j^*\}} j^*$}{$S_0=\{S_0\setminus\{j^*\} \}\cup\{i^*\}$}
 }
\Output{Set $S_0$  of cardinality $K$.}

\caption{Constructive heuristic for  the EWCCMVSF problem }\label{heuristic}
\end{algorithm}\DecMargin{1em}

\IncMargin{1em}
\begin{algorithm}
\SetKwData{Left}{left}\SetKwData{This}{this}\SetKwData{Up}{up}
\SetKwFunction{Union}{Union}\SetKwFunction{FindCompress}{FindCompress}
\SetKwInOut{Input}{input}\SetKwInOut{Output}{output}
\Input{A set $S_0$ from Algorithm 1, and a set $A=\left \{( \beta_{i} \sigma_f,\sigma_{\epsilon_i}^2), \, \forall  i\in S_0 \subset I\right \}$.}
\BlankLine
\Repeat{$j^* <<_{S_0\setminus \{j^*\}} i^*$}{
\For{$i\in S_0$}{
Calculate $obj(S_0)-obj(S\setminus\{i\})=\sigma_{\epsilon_i}^2+ \sigma^2_f\left (( \sum_{k\in S_0}\beta_k)^2 -(\sum_{k\in S_0:k\neq i}\beta_k)^2 \right)$
}\emph{}
 \emph{ Let $i^*=\arg_{i\in S_0} \max \{obj(S_0)-obj(S\setminus\{i\}) \}$, i.e., $i^*$ is the asset in $S_0\subset I$ with the biggest contribution in the objective function.}\\
  \lIf{$iobj(S_0) > \frac{K^2}{(K-1)^2} obj(S_0\setminus \{i^*\} )$}{$S_0=S_0\setminus\{i^*\} $, and $K=K-1$.}
 }
\Output{Set $S_0$  of cardinality $K$.}

\caption{Improving the solution of Algorithm 1}\label{improving}
\end{algorithm}\DecMargin{1em}

\section{Computational Results}

In this section we present the results obtained from the computational experience. We have generated several instances from the index tracking instances available at the OR-Library \citep{Beasley1990}. A full list of the test datasets in the OR-Library, for a single factor model, can be found in  \url{http://people.brunel.ac.uk/~mastjjb/jeb/orlib/indtrackinfo.html}. The instances selected are indtrack5,6,7 and 8, the biggest. These datasets have been used in several papers, see \citep{Beasley2003,Beasley2009,chang2000,Beasley2011}.
  Each dataset contains the weekly market price for a set of assets and the market index.  Additionally, we have considered for each dataset their four principal components in order to use these components as factors and evaluate the factor models presented in section 2.

The computational experiments were conducted on a PC with 2.9 gigahertz Intel Core i5 processor, 8gigabytes of RAM, and operating system OX. We use the optimization engine CPLEX v12.5.

We have divided the computational experience into three  parts. First, we compare the performance of the different models we have proposed for  factor models. Next, we repeat the computational experiment for a single factor model. Finally, we  have generated an ad hoc instance to take models and the algorithm   to the limit.

\subsection{Computational results for a  factor models.}

For each dataset considered, we have calculated their first four principal components, and then, the $\beta$ and $\sigma_{\epsilon_i}^2$ for each asset in these components (factors). Note that the use of principal components as  factors provide factors which are  uncorrelated. The computational experience is performed on the following four models: $CCMVFM$, $CCMVFM_{LA}$, $EWCCMVFM$ and $EWCCMVFM_{LA}$.  
Each dataset is solved for different  values of cardinality parameter $K$. Tables \ref{indtrack5_MF}-\ref{indtrack8_MF} show the computational results for the four models in each dataset (the caption of each table collects the dataset name, the market and the number of assets),  where the columns for each model and cardinality considered are as follows: $time$, elapsed time to obtain the optimal solution or the time limit of 3600 seconds; $obj$, solution value; $\%desv=100(obj(\cdot)-obj(CCMVFM)/obj(CCMVFM))$, deviation of the solution value obtained by the model from the solution of CCMVFM problem; $K$, the number of assets in the solution of the CCMVFM problem and the number of assets in the solution together with  the number of assets   that coincide with the solution of the problem CCMVFM;   $||w-w^*||_1$,  $L_1$ distance of the solution variables from the solution variables of CCMVFM problem; $SD=\sqrt{w^T\Sigma w}$, standard deviation of each solution, $\%desv$, \%deviation of $SD$ for each model respect the model ($CCMVFM$); and $SR$ the ratio return/risk ($w^TR/\sqrt{w^T\Sigma w}$) for each model. 

\emph{Quality evaluation of   $CCMVFM_{LA}$ solution}. If we focus  attention on Table \ref{indtrack8_MF} (biggest dataset considered), we can observe the very small elapsed time that is required and the goodness of the solution ($\%desv$) versus the one provided by the model ($CCMVFM$), it is not too-high. You can also observe  that CPLEX reaches the time limit (1 h.) in all the instances we have experimented with,  for different values of cardinality $K$, while the elapsed time of the $CCMVFM_{LA}$ problems is one or two orders of magnitude smaller than the CPLEX limit considered. For example, we should point out the instance for cardinality $K=10$, where the linear approximation model ($CCMVFM_{LA}$) obtains a solution with a deviation of 0.16\% from the ($CCMVFM$) model in approximately  20 seconds, and a  6\% better standard deviation than the standard deviation  of ($CCMVFM$) model. Despite the fact that the numbers of identical assets in both solutions are  only 5 of 10, with a $L_1$ distance of 0.88. These comments are also valid for the smaller datasets, see Tables \ref{indtrack5_MF}-\ref{indtrack7_MF}.

\emph{Quality evaluation of  $EWCCMVFM$ and  $EWCCMVFM_{LA}$ solutions}. First we can observe the very little elapsed time that is required to obtain the optimal solution of $EWCCMVFM$ and its approximation  $EWCCMVFM_{LA}$. In any instance more than  3 seconds is required.  The solutions of $EWCCMVFM$ and its approximation are very similar, therefore the  $EWCCMVFM_{LA}$ does not provide any advantages to justify  its use. In Table \ref{indtrack8_MF}, the deviation of  $EWCCMVFM$ solution from  $CCMVFM$ solution varies from 0.48\% to 7.55\%. This  difference  comes from  imposing  the equality weighted constraint on  the solution. Nevertheless, the equality weighted  solution provides better results, in some instances, when the standard deviation and the ratio 
 return/risk are evaluated. An exception occurs in Table  \ref{indtrack5_MF}, where setting the cardinality parameter $K$ to 20 or 30 forces us  to select a larger number of assets when no more than 13 are suitable.

The dimensions of each problem can be obtained from Table \ref{t:dimension}. Table \ref{t:dimension_indtrack8} shows the dimensions of the largest instance considered for each model, which  has been obtained from the indtract8.txt dataset;  this instance contains  $N=2151$ assets from the Russel 300 index and the parameter of cardinality fixed to 50. The number of segments in the linear approximations have been fixed to $|S_w|=4\cdot K +1=4\cdot 50 +1 = 201$, and  $|S_\beta|=  500$.  Although the problem $EWCCMVFM_{LA}$ for this instance has  more than four hundred thousand binary variables, CPLEX only needs 155 seconds to solve it.  

In summary, from the results obtained by the models, we can deduce from our preliminary computational experimentation that the solution values do not differ too-much. $CCMVFM$ problems require a high elapsed time,  while the rest of models are very fast, in fact the elapsed time  can be measured in a few seconds.

\subsection{Computational results for a single factor model.}

We next compare the performance of the models and algorithm we have proposed in section 3 for a single index factor. We have used the same data sets and the index included  in them. Additionally, we have replaced the model names with  their counterpart  names in a a  single factor model, and we have replaced also the approximation of $EWCCMVSF$, ($EWCCMVSF_{LA}$),  by the algorithms proposed in section 3. 

Tables  \ref{indtrack5_1F}-\ref{indtrack8_1F} show the same information as in Tables \ref{indtrack5_MF}-\ref{indtrack8_MF},  but for the  single factor model. 

We first discuss the small instances presented in Table \ref{indtrack5_1F}. Obviously, in the Small network the time differences  between the models are slight. In the instances with cardinality 20 and 30, since we have imposed these parameters of  cardinality, the  $EWCCMVSF$ model obtains a much worse solution with 20 and 30 asset when the optimal solution for $CCMVSF$ problem  is selected only 16 asset. In this sense, the algorithm 2 improves the solution by removing assets from the solution, obtaining as a result a maximum cardinality of 9 assets  for this instance.

 For the sake of simplicity we now discuss  only the biggest instance, Table \ref{indtrack8_1F}, but similar conclusions can be drawn from the other two sets of instances reported in tables \ref{indtrack6_1F} and \ref{indtrack7_1F}.  Our first observation is that the computing time for solving problem ($CCMVSF$) (i.e, the original problem by plain use of CPLEX) is high for all the instances (1 h in our experimentation is the allowed computing time). On the other hand, the linear approximation ($CCMVSF_{LA}$) requires only a few seconds to obtain a solution, while the deviation is only of 0.28\% in the worst case (instance with parameter of cardinality $K=30$), providing even better results in some instances than  the solution obtained by the $CCMVSF$ problem. Comparing now the results obtained from the the $EWCCMVSF$ model, the time spent on the $EWCCMVSF$ problem  in all the  instances is less than  three  seconds, this is a consequence of the Proposition 3 in section 3. Moreover, the quality of the solutions obtained from ($EWCCMV$) is high, with a deviation of 4.35\% in the worst instance and selecting 48 of 50 assets present in the solution of ($CCMVSF$) problem and a $L_1$ distance of 0.31. The good quality of solutions  is also observed in the standard deviation $SD$ a ratio $SR$ of them, all the solutions  being  close to each other.

From a practical point of view, we have evaluated the validity of the models presented ins this work. We believe that the models, especially the equality weighted models can be helpful to the practitioners to evaluate the best assets to consider and in a posterior analysis to apply other  more complex techniques .

Finally, it can be seen in tables \ref{indtrack5_1F}-\ref{indtrack8_1F} that when $K$ increases all the measures take similar values.

\subsection{An ad hoc instance}

In Order to test the models and the  algorithm in the case of a more difficult problem, we have built the following instance, called indtrack5678. For a single factor model we added  all the $\beta_i$ and $\sigma_{\epsilon_i}$ from the data sets indtrack5,  indtrack6, indtrack7 and indtrack8. This new dataset contains 4151 assets.

In  Figure 1 we plot one point for each asset $i\in I$, representing the systematic ($\beta_i$) and non-systematic ($\sigma_{\epsilon_i}^2$) risk for each of them. In the original dataset (figure on the left) the cloud of points is located around all the graph region. Ideally, one would like to have points near the intersection of axes which represent low risk (systematic and non-systematic). On the other hand,  located points not close to the intersection of the axes are dominated for the remaining  points, and hopefully  these points will not be present in the solution of  CCMVFM problem. This feature in the datasets make the instances more treatable, computationally speaking. 

We call $ad$ $hoc$ instance, the instance indtrack5678 where the assets have been sorted by the systematic risk $\beta_i$ value (from lowest to highest ones) and matching each $\beta_i$ value with the non-systematic risk ($\sigma_{\epsilon_i}^2$) sorted in reverse order.  This $ad$ $hoc$ instance  provides non dominated assets between them, non dominated in Pareto sense. The cloud of points for the ad-hoc instance (figure on the right) is structured, because all the points are non-dominated. The same consideration taken above is valid here, the best assets are located near to the intersection axes, but now these points are non-dominated among these. Therefore, this new instance is more difficult for the factor models problem than the previous example. In Table \ref{indtrack678} we report the results for this structured data set.

First of all, we can see that the algorithm does not provide the optimal solution in 3 of 6 instances. Another important feature in the results is 
that the $EWCCMV$ model obtains the best solution in 4 of the 6 instances,  exactly for the values of parameter $K$ equal to  5, 10, 20 and 40. The $CCMVSF$ model is the best for the rest of the  instances, but requiring one hour of computational time.  The solutions obtained suggest multiple alternative choice of assets. For example, in the instance with cardinality 30, the objective solution value of $CCMVSF$ and $EWCCMVSF$ are quite similar (only 0.01\% of deviation) but quality  speaking are very different, they only have 11   assets in common.

\section{Conclusions}

In this paper we have proposed and analyzed two alternative model to obtain the cardinality constrained minimum-variance portfolio via factor models. The intention in both models is to obtain a linear model in contrast with the quadratic factor model present in the literature. This goal is reached when the factors are uncorrelated. This  assumption is not  very  restrictive in the financial context. 

Regarding the comparison  of the models, the Equality Weighted cardinality constrained portfolio problem has provided the most promising results, obtaining goods solutions. In terms of computational time, all the instances require less than three second to solve them. On the other hand, the heuristic presented in this paper, when a single factor is considered, obtains the optimal solution in all the instances,  with  the exception of the ad-hoc instance generated.  Therefore, the EWCCMF model and the heuristic approach can be regarded as being superior to the classical factor models in terms of usability; it obtains high quality solutions with little   computational  time.

Furthermore, since the models presented in this work have provided good results, we also plan to extend these models, for example, when the objective is to minimise a trade-off function risk/return. 

%
%

\section*{Acknowledgments}

This work was partly supported by the  Spanish Ministry for Economy and Competitiveness, the State Research Agency  and the European Regional Development Fund  under grant MTM2016-79765-P (AEI/FEDER, UE).

\section*{Appendix.  Can be solved the EWCCMVSF problem in polynomial time?}

\begin{definition}  A matrix $M$ is a $Monge$ $matrix$ if for every pair of rows $i<j$ and for every pair of columns $k<l$ satisfies the $Monge$ $property$ 
\begin{align}
 M_{ik}+M_{jl}\leq M_{il}+M{jk}.
\label{Monge}
\end{align}
\end{definition}
\begin{definition}  A matrix $M$ is called an $inverse$ $Monge$ $matrix$ if it satisfies the inverse $Monge$ $property$ 
\begin{align}
 M_{ik}+M_{jl}\geq M_{il}+M{jk}, \quad  \text{for all} \quad  i<j, \, k<l.
\label{Monge}
\end{align}
\end{definition}
Note that  a symmetric Monge matrix is called a $Supnick$ matrix.

Monge matrices have many applications in combinatorial optimization problems, see \citep{pferschy1994,Woeginger2003,Burkard1996,Rudolf1995}. For example, the Traveling Salesman Problem (TSP) can be solved in linear time if the underlying distance matrix is a Monge matrix, see \citep{park1991}.

\begin{proposition} The underlying matrix in  ($EWCCMVSF$ (\ref{EWCCMVSF})) is an inverse Monge matrix.
\end{proposition}
\proof{Proof.} 
The objective function of the $EWCCMVSF$ model  can be expressed as
\[\min \quad \displaystyle  \frac{1}{K^2}      \sum_{i,j \in I}\left  ( \frac{\sigma_{\epsilon_i}^2+ \sigma_{\epsilon_j}^2}{2K}+ \sigma^2_f\beta_i \beta_j \right )\,  x_{i}x_{j}  \]
where $\sum_{i\in I}x_i=K$. The problem  (\ref{EWCCMVSF}) is given by the matrix  $M_{ij}=\left \{ \frac{\sigma_{\epsilon_i}^2+ \sigma_{\epsilon_j}^2}{2K}+ \sigma^2_f\beta_i \beta_j  \right \}_{ij}$. If we consider the ordered set $I$, i.e, $\beta_1\leq \beta_2, \dots \leq \beta_n$, is easy to prove that  $ M_{ik}+M_{jl}\geq M_{il}+M_{jk}$, for all $i<j$ y $k<l$.   
\begin{align}
&\left  ( \frac{\sigma_{\epsilon_i}^2+ \sigma_{\epsilon_k}^2}{2K}+ \sigma^2_f\beta_i \beta_k\right ) +\left  ( \frac{\sigma_{\epsilon_j}^2+ \sigma_{\epsilon_l}^2}{2K}+ \sigma^2_f\beta_j \beta_l\right ) \geq  
\left  ( \frac{\sigma_{\epsilon_i}^2+ \sigma_{\epsilon_l}^2}{2K}+ \sigma^2_f\beta_i \beta_l \right )+\left  ( \frac{\sigma_{\epsilon_j}^2+ \sigma_{\epsilon_k}^2}{2K}+ \sigma^2_f\beta_j \beta_k \right ) \nonumber \\
&\left  ( \frac{\sigma_{\epsilon_i}^2+ \sigma_{\epsilon_k}^2}{2K}+ \sigma^2_f\beta_i \beta_k\right ) +\left  ( \frac{\sigma_{\epsilon_j}^2+ \sigma_{\epsilon_l}^2}{2K}+ \sigma^2_f\beta_j \beta_l\right ) -  
\left  ( \frac{\sigma_{\epsilon_i}^2+ \sigma_{\epsilon_l}^2}{2K}+ \sigma^2_f\beta_i \beta_l \right )-\left  ( \frac{\sigma_{\epsilon_j}^2+ \sigma_{\epsilon_k}^2}{2K}+ \sigma^2_f\beta_j \beta_k \right )=\nonumber \\
=&  \sigma^2_f(\beta_i - \beta_j)(\beta_k-\beta_l) \geq 0 \nonumber
\end{align}
\Halmos
\endproof

Therefore, finding  the equality weighted cardinality constrained  portfolio for a single factor model is reduced to finding  the $K$ columns/rows in the matrix $\left \{\frac{\sigma_{\epsilon_i}^2+ \sigma_{\epsilon_i}^2}{2K}+ \sigma^2_f\beta_i \beta_j \right \}_{ij}$ with lower cost. An open question is as follows:

\begin{center}
Can $the$ $EWCCMVSF$ $problem$  $be$ $solved$ $in$ $polynomial$ $time$?
\end{center}

I did not find the answer to the above question and I suggest that   the reader  might attempt to answer  this.

\newpage

\begin{table}[htb]
\extrarowheight = -1.3ex
\centering
\begin{tabular}{l@{\hspace{1.1cm}}c@{\hspace{1.25cm}}c@{\hspace{1.25cm}}c} \hline
    Model          &   $Linear$   &  $Quadratic$   &  $Quadratic$  \\ 
    	    	&    pure $01$  &   pure $01$   &  $mixed$ $01$\\ \hline
   $CCMVFM$   &  -  &  -   &   $ UFM,CFM$   \\ 
    $CCMVFM_{LA}$  &  $UFM$  & $CFM$  &   - \\ 
   $EWCCMVFM$             & -   &  $ UFM,CFM$  &    -  \\
   $EWCCMVFM_{LA}$ &  $ UFM$ & $CFM$ & -     \\ \hline
   \multicolumn{4}{l}{UFM; Uncorrelated Factor  Models. CFM; Correlated Factor Models}\\
\end{tabular}
\caption{Classification Models}
\label{t:classification}
\end{table}

\begin{table}[htb]
\extrarowheight = -1.3ex
\centering
\begin{tabular}{l@{\hspace{1.5cm}}c@{\hspace{1.5cm}}r@{\hspace{1.5cm}}c} \hline
    Model &   $n01$   &  $nc$   & $m$ \\ \hline
   $CCMVFM$   &   $N$  &   $N$  &   $N+2$ \\ 
   $CCMVFM_{LA}$  &  $N\cdot |S_w|+ NF\cdot |S_{\beta}| $  & -  &   $N+2 NF +2$ \\ 
     $EWCCMVFM$             &    $N$   &   -  &    $1$  \\
   $EWCCMVFM_{LA}$   &    $N+NF\cdot |S_{\beta}| $   &  - &   $2 NF +2$  \\ \hline
   \multicolumn{4}{l}{$ |S_w|=4\cdot K +1$, and $ |S_{\beta}|=500$} \\
\end{tabular}
\caption{Dimension Models}
\label{t:dimension}
\end{table}

\begin{table}[htb]
\extrarowheight = -1.3ex
\footnotesize
\begin{tabular}{lllrrrrrrrrrrr} 	\hline	
$N=225$ &&&&	\multicolumn{5}{c}{Solution}	&&	\multicolumn{3}{c}{}			\\   
 \cline{1-3}  \cline{5-9}	\cline{11-13}
&& $Model$	&&	$time$	&	$obj$	&	$\%desv$	&	$K$			&$\displaystyle	||w-w^*||_1$	&&	$SD$	&	$\%desv$	&	$SR$	\\
  \cline{1-3}  \cline{5-9}	\cline{11-13}	
$K=5$ &&	$CCMVFM$		&&	0.11	&	0.01939	&	&	5	&	&	&	0.01971	&	&	0.00000	\\		
&&		$CCMVFM_{LA}$		&&	0.90	&	0.01941	&	0.10	\%	&	5	-	5	&	0.21	&	&	0.02024	&	2.73	\%	&	-0.00419	\\
&&	$EWCCMVFM$	&&	0.03	&	0.01967	&	1.37	\%	&	5	-	5	&	0.24	&	&	0.02006	&	1.78	\%	&	-0.00315	\\		
&&		$EWCCMVFM_{LA}$	&&	0.04	&	0.01968	&	1.40	\%	&	5	-	5	&	0.24	&	&	0.02006	&	1.78	\%	&	-0.00315	\\	
\hline
$K=10$&&	$CCMVFM$		&&	0.12	&	0.01886	&	&	10	&	&	&	0.01945	&	&	0.01633	\\								
&&		$CCMVFM_{LA}$		&&	0.82	&	0.01897	&	0.61	\%	&	10	-	8	&	0.39	&	&	0.01998	&	2.74	\%	&	0.00942	\\
&&	$EWCCMVFM$	&&	0.07	&	0.01967	&	4.30	\%	&	10	-	8	&	0.71	&	&	0.02029	&	4.30	\%	&	-0.00263	\\		
&&		$EWCCMVFM_{LA}$	&&	0.09	&	0.01967	&	4.32	\%	&	10	-	8	&	0.71	&	&	0.02029	&	4.30	\%	&	-0.00263	\\	
\hline
$K=20$ &&	$CCMVFM$		&&	0.06	&	0.01882	&	&	13	&	&	&	0.01938	&	&	0.01180	\\								
&&		$CCMVFM_{LA}$		&&	1.45	&	0.01887	&	0.26	\%	&	15	-	13	&	0.17	&	&	0.01970	&	1.62	\%	&	0.00855	\\
&&	$EWCCMVFM$	&&	0.31	&	0.02064	&	9.63	\%	&	20	-	13	&	0.98	&	&	0.02112	&	8.97	\%	&	-0.00513	\\		
&&		$EWCCMVFM_{LA}$	&&	0.20	&	0.02064	&	9.66	\%	&	20	-	13	&	0.98	&	&	0.02112	&	8.97	\%	&	-0.00513	\\	
\hline
$K=30$ &&	$CCMVFM$		&&	0.03	&	0.01882	&	&	13	&	&	&	0.01938	&	&	0.01180	\\								
&&		$CCMVFM_{LA}$		&&	1.66	&	0.01887	&	0.25	\%	&	15	-	13	&	0.17	&	&	0.01958	&	1.04	\%	&	0.00874	\\
&&	$EWCCMVFM$	&&	0.36	&	0.02170	&	15.31	\%	&	30	-	13	&	1.22	&	&	0.02200	&	13.52	\%	&	0.00376	\\		
&&		$EWCCMVFM_{LA}$	&&	0.20	&	0.02171	&	15.32	\%	&	30	-	13	&	1.22	&	&	0.02200	&	13.52	\%	&	0.00376	\\	 \hline
 \multicolumn{13}{l}{Time limit 3600 sec.}
\end{tabular}
 \caption{indtrack5.txt \quad Nikkei  225 index. \, N=225}  
 \label{indtrack5_MF}
\end{table}

\begin{table}[htb]
\extrarowheight = -1.3ex
\footnotesize
\begin{tabular}{lllrrrrrrrrrrr} 	\hline	
$N=457$ &&&&	\multicolumn{5}{c}{Solution}	&&	\multicolumn{3}{c}{}			\\   
 \cline{1-3}  \cline{5-9}	\cline{11-13}
&& $Model$	&&	$time$	&	$obj$	&	$\%desv$	&	$K$			&$\displaystyle	||w-w^*||_1$	&&	$SD$	&	$\%desv$	&	$SR$	\\
  \cline{1-3}  \cline{5-9}	\cline{11-13}	
$K=5$  &&	$CCMVFM$		&&	653.98	&	0.02030	&	&	5	&	&	&	0.02272	&	&	0.13508	\\								
&&		$CCMVFM_{LA}$		&&	4.90	&	0.02043	&	0.66	\%	&	5	-	3	&	1.15	&	&	0.02308	&	1.59	\%	&	0.13842	\\
&&	$EWCCMVFM$	&&	0.28	&	0.02174	&	7.09	\%	&	5	-	2	&	1.20	&	&	0.02240	&	-1.40	\%	&	0.14915	\\		
&&		$EWCCMVFM_{LA}$	&&	0.27	&	0.02175	&	7.14	\%	&	5	-	2	&	1.20	&	&	0.02240	&	-1.40	\%	&	0.14915	\\	
 \hline
$K=10$ &&	$CCMVFM$		&&	3600.00	&	0.01703	&	&	10	&	&	&	0.01830	&	&	0.13563	\\								
&&		$CCMVFM_{LA}$		&&	1.33	&	0.01712	&	0.51	\%	&	10	-	8	&	0.51	&	&	0.01948	&	6.45	\%	&	0.16445	\\
&&	$EWCCMVFM$	&&	1.23	&	0.01813	&	6.46	\%	&	10	-	6	&	0.80	&	&	0.01914	&	4.54	\%	&	0.14137	\\		
&&		$EWCCMVFM_{LA}$	&&	0.30	&	0.01814	&	6.48	\%	&	10	-	5	&	1.00	&	&	0.01942	&	6.08	\%	&	0.15028	\\	 
 \hline
$K=20$&&	$CCMVFM$		&&	3600.00&	0.0155	&	&	20	&	&	&	0.01681	&	&	0.16360	\\								
&&		$CCMVFM_{LA}$		&&	3.06	&	0.01553	&	0.17	\%	&	20	-	18	&	0.29	&	&	0.01769	&	5.26	\%	&	0.16235	\\
&&	$EWCCMVFM$	&&	0.76	&	0.01603	&	3.42	\%	&	20	-	16	&	0.57	&	&	0.01790	&	6.50	\%	&	0.14760	\\		
&&		$EWCCMVFM_{LA}$	&&	0.13	&	0.01604	&	3.49	\%	&	20	-	16	&	0.57	&	&	0.01790	&	6.50	\%	&	0.14760	\\	
 \hline
$K=30$ &&	$CCMVFM$		&&	2.93	&	0.01510	&	&	30	&	&	&	0.01718	&	&	0.15862	\\								
&&		$CCMVFM_{LA}$		&&	5.41	&	0.01525	&	0.95	\%	&	30	-	30	&	0.22	&	&	0.01757	&	2.31	\%	&	0.15265	\\
&&	$EWCCMVFM$	&&	0.26	&	0.01564	&	3.53	\%	&	30	-	26	&	0.48	&	&	0.01782	&	3.73	\%	&	0.17109	\\		
&&		$EWCCMVFM_{LA}$	&&	0.18	&	0.01565	&	3.65	\%	&	30	-	25	&	0.54	&	&	0.01781	&	3.70	\%	&	0.16706	\\	
 \hline
 \multicolumn{13}{l}{Time limit 3600 sec.}
\end{tabular}
 \caption{indtrack6.txt \quad S\&P 500 index. \, N=457}  
 \label{indtrack6_MF}
\end{table}

\begin{table}[htb]
\extrarowheight = -1.3ex
\footnotesize
\begin{tabular}{lllrrrrrrrrrrr} 	\hline	
$N=1318$ &&&&	\multicolumn{5}{c}{Solution}	&&	\multicolumn{3}{c}{}			\\   
 \cline{1-3}  \cline{5-9}	\cline{11-13}
&& $Model$	&&	$time$	&	$obj$	&	$\%desv$	&	$K$			&$\displaystyle	||w-w^*||_1$	&&	$SD$	&	$\%desv$	&	$SR$	\\
  \cline{1-3}  \cline{5-9}	\cline{11-13}	
$K=5$  &&	$CCMVFM$		&&	3600.00&	0.01134	&	&	5	&	&	&	0.01298	&	&	0.10390	\\								
&&		$CCMVFM_{LA}$		&&	33.96	&	0.01205	&	6.24	\%	&	5	-	3	&	0.85	&	&	0.01446	&	11.45	\%	&	0.09846	\\
&&	$EWCCMVFM$	&&	0.87	&	0.01149	&	1.30	\%	&	5	-	4	&	0.46	&	&	0.01235	&	-4.83	\%	&	0.10731	\\		
&&		$EWCCMVFM_{LA}$	&&	0.16	&	0.01150	&	1.42	\%	&	5	-	5	&	0.12	&	&	0.01299	&	0.10	\%	&	0.10520	\\	
 \hline
$K=10$&&	$CCMVFM$		&&	3600.00	&	0.00904	&	&	10	&	&	&	0.01124	&	&	0.11999	\\								
&&		$CCMVFM_{LA}$		&&	13.63	&	0.00931	&	2.98	\%	&	10	-	8	&	0.37	&	&	0.01128	&	0.33	\%	&	0.12247	\\
&&	$EWCCMVFM$	&&	0.58	&	0.00918	&	1.49	\%	&	10	-	8	&	0.48	&	&	0.01102	&	-1.94	\%	&	0.15623	\\		
&&		$EWCCMVFM_{LA}$	&&	0.15	&	0.00924	&	2.14	\%	&	10	-	9	&	0.29	&	&	0.01068	&	-5.01	\%	&	0.13644	\\	
 \hline
$K=20$&&	$CCMVFM$		&&	3600.00	&	0.00729	&	&	20	&	&	&	0.00982	&	&	0.13600	\\								
&&		$CCMVFM_{LA}$		&&	45.96	&	0.00765	&	4.89	\%	&	20	-	15	&	0.58	&	&	0.00968	&	-1.45	\%	&	0.17680	\\
&&	$EWCCMVFM$	&&	0.55	&	0.00741	&	1.54	\%	&	20	-	18	&	0.31	&	&	0.00978	&	-0.47	\%	&	0.12624	\\		
&&		$EWCCMVFM_{LA}$	&&	0.15	&	0.00753	&	3.30	\%	&	20	-	19	&	0.26	&	&	0.01018	&	3.61	\%	&	0.13306	\\	
 \hline
$K=30$ &&	$CCMVFM$		&&	3600.00	&	0.00656	&	&	30	&	&	&	0.00914	&	&	0.16385	\\								
&&		$CCMVFM_{LA}$		&&	155.05	&	0.00697	&	6.36	\%	&	30	-	22	&	0.61	&	&	0.00943	&	3.13	\%	&	0.21373	\\
&&	$EWCCMVFM$	&&	0.89	&	0.00676	&	3.11	\%	&	30	-	25	&	0.40	&	&	0.00950	&	3.93	\%	&	0.15389	\\		
&&		$EWCCMVFM_{LA}$	&&	0.19	&	0.00686	&	4.65	\%	&	30	-	25	&	0.41	&	&	0.00923	&	0.94	\%	&	0.15019	\\	
 \hline
$K=40$&&	$CCMVFM$		&&	3600.00	&	0.00621	&	&	40	&	&	&	0.00849	&	&	0.19668	\\								
&&		$CCMVFM_{LA}$		&&	195.53	&	0.00660	&	6.26	\%	&	40	-	31	&	0.50	&	&	0.00895	&	5.34	\%	&	0.22163	\\
&&	$EWCCMVFM$	&&	0.75	&	0.00650	&	4.71	\%	&	40	-	33	&	0.48	&	&	0.00915	&	7.75	\%	&	0.16893	\\		
&&		$EWCCMVFM_{LA}$	&&	0.33	&	0.00659	&	6.11	\%	&	40	-	34	&	0.41	&	&	0.00874	&	2.92	\%	&	0.18297	\\	
\hline
 $K=50$&&	$CCMVFM$		&&	3600.00	&	0.00603	&	&	50	&	&	&	0.00867	&	&	0.19824	\\								
&&		$CCMVFM_{LA}$		&&	183.74	&	0.00642	&	6.50	\%	&	50	-	41	&	0.40	&	&	0.00863	&	-0.36	\%	&	0.19161	\\
&&	$EWCCMVFM$	&&	0.60	&	0.00643	&	6.65	\%	&	50	-	42	&	0.53	&	&	0.00903	&	4.15	\%	&	0.20117	\\		
&&		$EWCCMVFM_{LA}$	&&	0.13	&	0.00654	&	8.37	\%	&	50	-	42	&	0.52	&	&	0.00911	&	5.11	\%	&	0.19687	\\	\hline
 \multicolumn{13}{l}{Time limit 3600 sec.}
\end{tabular}
 \caption{indtrack7.txt \quad Russel 2000 index. \, N=1318}  	
 \label{indtrack7_MF}
\end{table}

\begin{table}[htb]
\extrarowheight = -1.3ex
\footnotesize
\begin{tabular}{lllrrrrrrrrrrr} 	\hline	
$N=2151$ &&&&	\multicolumn{5}{c}{Solution}	&&	\multicolumn{3}{c}{}			\\   
 \cline{1-3}  \cline{5-9}	\cline{11-13}
&& $Model$	&&	$time$	&	$obj$	&	$\%desv$	&	$K$			&$\displaystyle	||w-w^*||_1$	&&	$SD$	&	$\%desv$	&	$SR$	\\
  \cline{1-3}  \cline{5-9}	\cline{11-13}	
$K=5$  &&	$CCMVFM$		&&	3600.00	&	0.01173	&	&	5	&	&	&	0.01141	&	&	0.10210	\\								
&&		$CCMVFM_{LA}$		&&	115.95	&	0.01207	&	2.91	\%	&	5	-	4	&	0.76	&	&	0.01372	&	20.24	\%	&	0.10374	\\
&&	$EWCCMVFM$	&&	1.49	&	0.01190	&	1.47	\%	&	5	-	3	&	0.82	&	&	0.01197	&	4.84	\%	&	0.07009	\\		
&&		$EWCCMVFM_{LA}$	&&	0.43	&	0.01192	&	1.63	\%	&	5	-	3	&	0.82	&	&	0.01266	&	10.95	\%	&	0.06949	\\	
 \hline
$K=10$ &&	$CCMVFM$		&&	3600.00	&	0.00927	&	&	10	&	&	&	0.01127	&	&	0.11764	\\								
&&		$CCMVFM_{LA}$		&&	19.46	&	0.00928	&	0.16	\%	&	10	-	5	&	0.88	&	&	0.01058	&	-6.10	\%	&	0.10693	\\
&&	$EWCCMVFM$	&&	0.98	&	0.00931	&	0.48	\%	&	10	-	6	&	0.83	&	&	0.00999	&	-11.35	\%	&	0.14778	\\		
&&		$EWCCMVFM_{LA}$	&&	0.30	&	0.00940	&	1.39	\%	&	10	-	5	&	1.03	&	&	0.01008	&	-10.52	\%	&	0.17230	\\	
 \hline
$K=20$ &&	$CCMVFM$		&&	3600.00	&	0.00720	&	&	20	&	&	&	0.00968	&	&	0.13191	\\								
&&		$CCMVFM_{LA}$		&&	50.84	&	0.00733	&	1.82	\%	&	20	-	15	&	0.51	&	&	0.00936	&	-3.27	\%	&	0.16734	\\
&&	$EWCCMVFM$	&&	1.68	&	0.00753	&	4.60	\%	&	20	-	13	&	0.76	&	&	0.00961	&	-0.78	\%	&	0.15225	\\		
&&		$EWCCMVFM_{LA}$	&&	0.37	&	0.00751	&	4.23	\%	&	20	-	14	&	0.67	&	&	0.00949	&	-1.99	\%	&	0.15928	\\	
 \hline
$K=30$&&	$CCMVFM$		&&	3600.00	&	0.00644	&	&	30	&	&	&	0.00927	&	&	0.15817	\\								
&&		$CCMVFM_{LA}$		&&	102.04	&	0.00655	&	1.71	\%	&	30	-	21	&	0.54	&	&	0.00881	&	-4.95	\%	&	0.17737	\\
&&	$EWCCMVFM$	&&	1.66	&	0.00679	&	5.54	\%	&	30	-	21	&	0.68	&	&	0.00933	&	0.71	\%	&	0.17964	\\		
&&		$EWCCMVFM_{LA}$	&&	0.71	&	0.00674	&	4.75	\%	&	30	-	20	&	0.73	&	&	0.00909	&	-1.86	\%	&	0.17367	\\	
 \hline
$K=40$ &&	$CCMVFM$		&&	3600.00	&	0.00602	&	&	40	&	&	&	0.00926	&	&	0.17618	\\								
&&		$CCMVFM_{LA}$		&&	138.7	&	0.00615	&	2.19	\%	&	40	-	32	&	0.41	&	&	0.00869	&	-6.14	\%	&	0.18834	\\
&&	$EWCCMVFM$	&&	1.83	&	0.00639	&	6.17	\%	&	40	-	30	&	0.65	&	&	0.00951	&	2.78	\%	&	0.17404	\\		
&&		$EWCCMVFM_{LA}$	&&	0.68	&	0.00635	&	5.50	\%	&	40	-	32	&	0.56	&	&	0.00903	&	-2.49	\%	&	0.18674	\\	
 \hline
$K=50$&&	$CCMVFM$		&&	3600.00	&	0.00579	&	&	50	&	&	&	0.00976	&	&	0.17368	\\								
&&		$CCMVFM_{LA}$		&&	155.21	&	0.00593	&	2.44	\%	&	50	-	43	&	0.31	&	&	0.00905	&	-7.29	\%	&	0.19553	\\
&&	$EWCCMVFM$	&&	2.74	&	0.00622	&	7.55	\%	&	50	-	38	&	0.63	&	&	0.00916	&	-6.19	\%	&	0.19066	\\		
&&		$EWCCMVFM_{LA}$	&&	0.48	&	0.00622	&	7.49	\%	&	50	-	38	&	0.63	&	&	0.00925	&	-5.22	\%	&	0.18851	\\	
\hline
 \multicolumn{13}{l}{Time limit 3600 sec.}
\end{tabular}
 \caption{indtrack8.txt \quad Russel 3000 index. \, N=2151}  	
 \label{indtrack8_MF}
\end{table}

\begin{table}[htb]
\extrarowheight = -1.3ex
\centering
\begin{tabular}{l@{\hspace{1.5cm}}r@{\hspace{1.5cm}}r@{\hspace{1.5cm}}r} \hline
    $Model$ &   $n01$   &  $nc$   & $m$ \\ \hline
   $CCMVFM$   &   $2151$  &   $2151$  &   $2153$ \\ 
   $CCMVFM_{LA}$  &  $434351 $  & - &   $2161$ \\ 
     $EWCCMVFM$             &    $2151$   &   -  &    $1$  \\
   $EWCCMVFM_{LA}$   &    $4151 $   &  - &   $10$  \\ \hline
\end{tabular}
\caption{Dimension Model for $N=2151$ and $K=50$.}
\label{t:dimension_indtrack8}
\end{table}

\begin{table}[htb]
\extrarowheight = -1.3ex
\footnotesize
\begin{tabular}{lllrrrrrrrrrrr} 	\hline	
$N=225$ &&&&	\multicolumn{5}{c}{Solution}	&&	\multicolumn{3}{c}{}			\\   
 \cline{1-3}  \cline{5-9}	\cline{11-13}
&& $Model$	&&	$time$	&	$obj$	&	$\%desv$	&	$K$			&$\displaystyle	||w-w^*||_1$	&&	$SD$	&	$\%desv$	&	$SR$	\\
  \cline{1-3}  \cline{5-9}	\cline{11-13}	
$K=5$ &&	$CCMVSF$		&&	0.26	&	0.01820	&	&	5	&	&	&	0.01893	&	&	0.03109	\\								
&&	$CCMVSF_{LA}$	&&	0.13	&	0.01824	&	0.21	\%	&	5	-	4	&	0.39	&	&	0.0193	&	1.92	\%	&	0.03648	\\
&&	$EWCCMVSF$	&&	0.02	&	0.01831	&	0.61	\%	&	5	-	5	&	0.15	&	&	0.0193	&	1.94	\%	&	0.02940	\\		
&&	Alg.	(\ref{heuristic}) + Alg. (\ref{improving}) 		&&	0.00	&	0.01831	&	0.61	\%	&	5	-	5	&	0.15	&	&	0.0193	&	1.94	\%	&	0.02940	\\	
\hline
$K=10$&&	$CCMVSF$		&&	0.08	&	0.01736	&	&	10	&	&	&	0.01794	&	&	0.04348	\\								
&&	$CCMVSF_{LA}$			&&	0.17	&	0.01738	&	0.14	\%	&	10	-	10	&	0.10	&	&	0.01825	&	1.72	\%	&	0.04162	\\
&&	$EWCCMVSF$	&&	0.02	&	0.01766	&	1.71	\%	&	10	-	10	&	0.32	&	&	0.01833	&	2.15	\%	&	0.05960	\\		
&&	Alg.	(\ref{heuristic}) + Alg. (\ref{improving}) 		&&	0.00	&	0.01763	&	1.55	\%	&	9	-	9	&	0.31	&	&	0.01809	&	0.85	\%	&	0.04647	\\	
\hline
$K=20$ &&	$CCMVSF$		&&	0.03	&	0.01730	&	&	16	&	&	&	0.01816	&	&	0.04792	\\								
&&	$CCMVSF_{LA}$			&&	0.34	&	0.01732	&	0.09	\%	&	17	-	16	&	0.08	&	&	0.01831	&	0.84	\%	&	0.05180	\\
&&	$EWCCMVSF$	&&	0.02	&	0.01836	&	6.14	\%	&	20	-	16	&	0.86	&	&	0.01936	&	6.58	\%	&	0.06977	\\		
&&	Alg.	(\ref{heuristic}) + Alg. (\ref{improving}) 		&&	0.00	&	0.01763	&	1.90	\%	&	9	-	9	&	0.41	&	&	0.01809	&	-0.37	\%	&	0.04647	\\	
\hline
$K=30$ &&	$CCMVSF$		&&	0.03	&	0.01730	&	&	16	&	&	&	0.01816	&	&	0.04792	\\								
&&	$CCMVSF_{LA}$			&&	0.39	&	0.01732	&	0.09	\%	&	19	-	16	&	0.09	&	&	0.01826	&	0.55	\%	&	0.04999	\\
&&	$EWCCMVSF$	&&	0.02	&	0.01931	&	11.63	\%	&	30	-	16	&	1.16	&	&	0.02027	&	11.60	\%	&	0.05360	\\		
&&	Alg.	(\ref{heuristic}) + Alg. (\ref{improving}) 		&&	0.00	&	0.01763	&	1.90	\%	&	9	-	9	&	0.41	&	&	0.01809	&	-0.37	\%	&	0.04647	\\	
 \hline
 \multicolumn{13}{l}{Time limit 3600 sec.}
\end{tabular}
 \caption{indtrack5.txt \quad Nikkei  225 index. \, N=225}  
\label{indtrack5_1F}
 \end{table}

\begin{table}[htb]
\extrarowheight = -1.3ex
\footnotesize
\begin{tabular}{lllrrrrrrrrrrr} 	\hline	
$N=457$ &&&&	\multicolumn{5}{c}{Solution}	&&	\multicolumn{3}{c}{}			\\   
 \cline{1-3}  \cline{5-9}	\cline{11-13}
&& $Model$	&&	$time$	&	$obj$	&	$\%desv$	&	$K$			&$\displaystyle	||w-w^*||_1$	&&	$SD$	&	$\%desv$	&	$SR$	\\
  \cline{1-3}  \cline{5-9}	\cline{11-13}	
$K=5$  &&	$CCMVSF$		&&	1402.49	&	0.01299	&	&	5	&	&	&	0.02198	&	&	0.06554	\\								
&&	$CCMVSF_{LA}$			&&	0.23	&	0.01301	&	0.17	\%	&	5	-	5	&	0.05	&	&	0.02190	&	-0.37	\%	&	0.06486	\\
&&	$EWCCMVSF$	&&	0.07	&	0.01313	&	1.06	\%	&	5	-	5	&	0.13	&	&	0.02252	&	2.49	\%	&	0.06470	\\		
&&	Alg.	(\ref{heuristic}) + Alg. (\ref{improving}) 		&&	0.00	&	0.01313	&	1.06	\%	&	5	-	5	&	0.13	&	&	0.02252	&	2.49	\%	&	0.06470	\\	
 \hline
$K=10$ && $CCMVSF$			&&  3600.00  &  0.00985  &  		   &  10    &  		   & 		   &  0.02185  & 		   &   0.07077 \\ 
&& $CCMVSF_{LA}$  &&     0.33  &  0.00986  &     0.14 \%  &  10 - 9   &     0.22   &    &  0.02342  &      7.21 \%  &   0.06589 \\ 
&& $EWCCMVSF$  				&&     1.15  &  0.00992  &     0.77 \%  &  10 - 10   &     0.10   &    &  0.02227  &      1.91 \%  &   0.07036 \\ 
&& Alg.	(\ref{heuristic}) + Alg. (\ref{improving})    		&&     0.00  &  0.00992  &     0.77 \%  &  10 - 10   &     0.10   &    &  0.02227  &      1.91 \%  &   0.07036 \\ 
 \hline
$K=20$ &&	$CCMVSF$		&&	3600.00	&	0.00775	&	&	20	&	&	&	0.02085	&	&	0.07614	\\								
&&	$CCMVSF_{LA}$			&&	0.56	&	0.00776	&	0.13	\%	&	20	-	19	&	0.11	&	&	0.02061	&	-1.16	\%	&	0.07771	\\
&&	$EWCCMVSF$	&&	0.95	&	0.00783	&	1.01	\%	&	20	-	19	&	0.20	&	&	0.02049	&	-1.73	\%	&	0.07752	\\		
&&	Alg.	(\ref{heuristic}) + Alg. (\ref{improving}) 		&&	0.00	&	0.00783	&	1.01	\%	&	20	-	19	&	0.20	&	&	0.02049	&	-1.73	\%	&	0.07752	\\	
 \hline
$K=30$ &&	$CCMVSF$		&&	3600.00	&	0.00703	&	&	30	&	&	&	0.02024	&	&	0.07604	\\								
&&	$CCMVSF_{LA}$			&&	0.74	&	0.00704	&	0.12	\%	&	30	-	29	&	0.08	&	&	0.02099	&	3.73	\%	&	0.07390	\\
&&	$EWCCMVSF$	&&	0.74	&	0.00722	&	2.69	\%	&	30	-	29	&	0.24	&	&	0.02020	&	-0.19	\%	&	0.07605	\\		
&&	Alg.	(\ref{heuristic}) + Alg. (\ref{improving}) 		&&	0.00	&	0.00722	&	2.69	\%	&	30	-	29	&	0.24	&	&	0.02020	&	-0.19	\%	&	0.07605	\\	
 \hline
 \multicolumn{13}{l}{Time limit 3600 sec.}
\end{tabular}
 \caption{indtrack6.txt \quad S\&P 500 index. \, N=457}  
\label{indtrack6_1F}
 \end{table}

\begin{table}[htb]
\extrarowheight = -1.3ex
\footnotesize
\begin{tabular}{lllrrrrrrrrrrr} 	\hline	
$N=1318$ &&&&	\multicolumn{5}{c}{Solution}	&&	\multicolumn{3}{c}{}			\\   
 \cline{1-3}  \cline{5-9}	\cline{11-13}
&& $Model$	&&	$time$	&	$obj$	&	$\%desv$	&	$K$			&$\displaystyle	||w-w^*||_1$	&&	$SD$	&	$\%desv$	&	$SR$	\\
  \cline{1-3}  \cline{5-9}	\cline{11-13}	
$K=5$  &&	$CCMVSF$		&&	3600.00	&	0.01104	&	&	5	&	&	&	0.01221	&	&	0.10533	\\								
&&	$CCMVSF_{LA}$			&&	0.52	&	0.01105	&	0.05	\%	&	5	-	5	&	0.03	&	&	0.01247	&	2.13	\%	&	0.10567	\\
&&	$EWCCMVSF$	&&	0.67	&	0.01111	&	0.64	\%	&	5	-	5	&	0.10	&	&	0.01220	&	-0.14	\%	&	0.10625	\\		
&&	Alg.	(\ref{heuristic}) + Alg. (\ref{improving}) 		&&	0.00	&	0.01111	&	0.64	\%	&	5	-	5	&	0.10	&	&	0.01220	&	-0.14	\%	&	0.10625	\\	
 \hline
$K=10$ &&	$CCMVSF$		&&	3600.00	&	0.00867	&	&	10	&	&	&	0.01065	&	&	0.11182	\\								
&&	$CCMVSF_{LA}$			&&	1.19	&	0.00866	&	-0.15	\%	&	10	-	8	&	0.39	&	&	0.01069	&	0.38	\%	&	0.12050	\\
&&	$EWCCMVSF$	&&	0.99	&	0.00873	&	0.71	\%	&	10	-	8	&	0.44	&	&	0.01110	&	4.23	\%	&	0.12488	\\		
&&	Alg.	(\ref{heuristic}) + Alg. (\ref{improving}) 		&&	0.00	&	0.00873	&	0.71	\%	&	10	-	8	&	0.44	&	&	0.01110	&	4.23	\%	&	0.12488	\\	
 \hline
$K=20$ &&	$CCMVSF$		&&	3600.00	&	0.00690	&	&	20	&	&	&	0.00931	&	&	0.15070	\\								
&&	$CCMVSF_{LA}$			&&	3.14	&	0.00692	&	0.39	\%	&	20	-	19	&	0.14	&	&	0.00931	&	0.02	\%	&	0.15643	\\
&&	$EWCCMVSF$	&&	0.56	&	0.00702	&	1.80	\%	&	20	-	20	&	0.17	&	&	0.00941	&	1.07	\%	&	0.14976	\\		
&&	Alg.	(\ref{heuristic}) + Alg. (\ref{improving}) 		&&	0.01	&	0.00702	&	1.80	\%	&	20	-	20	&	0.17	&	&	0.00941	&	1.07	\%	&	0.14976	\\	
 \hline
$K=30$ &&	$CCMVSF$		&&	3600.00	&	0.00616	&	&	30	&	&	&	0.00880	&	&	0.17744	\\								
&&	$CCMVSF_{LA}$			&&	4.41	&	0.00619	&	0.46	\%	&	30	-	28	&	0.13	&	&	0.00897	&	1.96	\%	&	0.17802	\\
&&	$EWCCMVSF$	&&	0.45	&	0.00627	&	1.82	\%	&	30	-	28	&	0.27	&	&	0.00902	&	2.48	\%	&	0.17907	\\		
&&	Alg.	(\ref{heuristic}) + Alg. (\ref{improving}) 		&&	0.01	&	0.00627	&	1.82	\%	&	30	-	28	&	0.27	&	&	0.00902	&	2.48	\%	&	0.17907	\\	
 \hline
$K=40$  &&	$CCMVSF$		&&	3600.00	&	0.00577	&	&	40	&	&	&	0.00881	&	&	0.19180	\\								
&&	$CCMVSF_{LA}$			&&	5.17	&	0.00579	&	0.35	\%	&	40	-	38	&	0.13	&	&	0.00899	&	2.07	\%	&	0.18897	\\
&&	$EWCCMVSF$	&&	0.46	&	0.00596	&	3.29	\%	&	40	-	37	&	0.30	&	&	0.00888	&	0.84	\%	&	0.18926	\\		
&&	Alg.	(\ref{heuristic}) + Alg. (\ref{improving}) 		&&	0.01	&	0.00596	&	3.29	\%	&	40	-	37	&	0.30	&	&	0.00888	&	0.84	\%	&	0.18926	\\	
 \hline
$K=50$  && $CCMVSF$		&&  3600.00  &  0.00554  &  		   &  50    &  		   & 		   &  0.00865  & 		   &   0.20516 \\ 
&& $CCMVSF_{LA}$   &&     5.86  &  0.00557  &     0.58 \%  &  50 - 47   &     0.15   &    &  0.00883  &      2.12 \%  &   0.21575 \\ 
&& $EWCCMVSF$  				&&     0.41  &  0.00578  &     4.35 \%  &  50 - 48   &     0.31   &    &  0.00906  &      4.81 \%  &   0.19987 \\ 
&& Alg.	(\ref{heuristic}) + Alg. (\ref{improving})    		&&     0.01  &  0.00578  &     4.35 \%  &  50 - 48   &     0.31   &    &  0.00906  &      4.81 \%  &   0.19987 \\ \hline
 \multicolumn{13}{l}{Time limit 3600 sec.}
\end{tabular}
 \caption{indtrack7.txt \quad Russel 2000 index. \, N=1318}  	
\label{indtrack7_1F}
 \end{table}

\begin{table}[htb]
\extrarowheight = -1.3ex
\footnotesize
\begin{tabular}{lllrrrrrrrrrrr} 	\hline	
$N=2151$ &&&&	\multicolumn{5}{c}{Solution}	&&	\multicolumn{3}{c}{
}			\\   
 \cline{1-3}  \cline{5-9}	\cline{11-13}
&& $Model$	&&	$time$	&	$obj$	&	$\%desv$	&	$K$			&$\displaystyle	||w-w^*||_1$	&&	$SD$	&	$\%desv$	&	$SR$	\\
  \cline{1-3}  \cline{5-9}	\cline{11-13}	
$K=5$  &&	$CCMVSF$		&&	3600.00	&	0.01048	&	&	5	&	&	&	0.01193	&	&	0.11668	\\								
&&	$CCMVSF_{LA}$			&&	0.92	&	0.01049	&	0.11	\%	&	5	-	5	&	0.06	&	&	0.01226	&	2.78	\%	&	0.11645	\\
&&	$EWCCMVSF$	&&	0.92	&	0.01057	&	0.89	\%	&	5	-	5	&	0.13	&	&	0.01192	&	-0.06	\%	&	0.11939	\\		
&&	Alg.	(\ref{heuristic}) + Alg. (\ref{improving}) 		&&	0.01	&	0.01057	&	0.89	\%	&	5	-	5	&	0.13	&	&	0.01192	&	-0.06	\%	&	0.11939	\\	
 \hline
$K=10$ &&	$CCMVSF$		&&	3600.00 &	0.00833	&	&	10	&	&	&	0.01172	&	&	0.11569	\\								
&&	$CCMVSF_{LA}$			&&	1.78	&	0.00818	&	-1.75	\%	&	10	-	6	&	0.74	&	&	0.01155	&	-1.4	\%	&	0.11382	\\
&&	$EWCCMVSF$	&&	0.36	&	0.00828	&	-0.55	\%	&	10	-	7	&	0.64	&	&	0.01167	&	-0.39	\%	&	0.10813	\\		
&&	Alg.	(\ref{heuristic}) + Alg. (\ref{improving}) 		&&	0.01	&	0.00828	&	-0.55	\%	&	10	-	7	&	0.64	&	&	0.01167	&	-0.39	\%	&	0.10813	\\	
 \hline
$K=20$ &&	$CCMVSF$		&&	3600.00	&	0.00635	&	&	20	&	&	&	0.01187	&	&	0.11308	\\								
&&	$CCMVSF_{LA}$			&&	4.91	&	0.00636	&	0.16	\%	&	20	-	19	&	0.13	&	&	0.01146	&	-3.39	\%	&	0.1191	\\
&&	$EWCCMVSF$	&&	0.87	&	0.00645	&	1.52	\%	&	20	-	18	&	0.32	&	&	0.01265	&	6.58	\%	&	0.10659	\\		
&&	Alg.	(\ref{heuristic}) + Alg. (\ref{improving}) 		&&	0.04	&	0.00645	&	1.52	\%	&	20	-	18	&	0.32	&	&	0.01265	&	6.58	\%	&	0.10659	\\	
 \hline
$K=30$ &&	$CCMVSF$		&&	3600.00	&	0.00555	&	&	30	&	&	&	0.01226	&	&	0.11084	\\								
&&	$CCMVSF_{LA}$			&&	14.15	&	0.00556	&	0,28	\%	&	30	-	29	&	0,1	&	&	0,01233	&	0,62	\%	&	0,10308	\\
&&	$EWCCMVSF$	&&	0,51	&	0,00563	&	1,36	\%	&	30	-	29	&	0,19	&	&	0,01261	&	2,89	\%	&	0,10829	\\		
&&	Alg.	(\ref{heuristic}) + Alg. (\ref{improving}) 		&&	0,01	&	0,00563	&	1,36	\%	&	30	-	29	&	0,19	&	&	0,01261	&	2,89	\%	&	0,10829	\\	
\hline
$K=40$ &&	$CCMVSF$		&&	3600.00	&	0,00512	&	&	40	&	&	&	0,01237	&	&	0,12275	\\								
&&	$CCMVSF_{LA}$			&&	26,58	&	0,00512	&	-0,04	\%	&	40	-	37	&	0,19	&	&	0,01283	&	3,72	\%	&	0,11696	\\
&&	$EWCCMVSF$	&&	3,19	&	0,00519	&	1,39	\%	&	40	-	34	&	0,41	&	&	0,01281	&	3,56	\%	&	0,10080	\\		
&&	Alg.	(\ref{heuristic}) + Alg. (\ref{improving}) 		&&	0,04	&	0,00519	&	1,39	\%	&	40	-	34	&	0,41	&	&	0,01281	&	3,56	\%	&	0.10080	\\	
 \hline
 $K=50$ && $CCMVSF$ 			&& 3600.00  &  0.00480  &  		   &  50    &  		   & 		   &  0.01232  & 		   &   0.12428 \\ 
&& $CCMVSF_{LA}$   &&    12.50  &  0.00481  &     0.21 \%  &  50 - 48   &     0.10   &    &  0.01239  &      0.58 \%  &   0.12879 \\ 
&& $EWCCMVSF$  				&&     0.52  &  0.00491  &     2.22 \%  &  50 - 47   &     0.27   &    &  0.01224  &     -0.63 \%  &   0.13159 \\ 
&& Alg.	(\ref{heuristic}) + Alg. (\ref{improving})    		&&     0.01  &  0.00491  &     2.22 \%  &  50 - 47   &     0.27   &    &  0.01224  &     -0.63 \%  &   0.13159 \\  \hline
 \multicolumn{13}{l}{Time limit 3600 sec.}
\end{tabular}
 \caption{indtrack8.txt \quad Russel 3000 index. \, N=2151}  	
\label{indtrack8_1F}
 \end{table}

\begin{figure}[htb]
\label{adhoc_fig}
\includegraphics[width=0.4\linewidth]{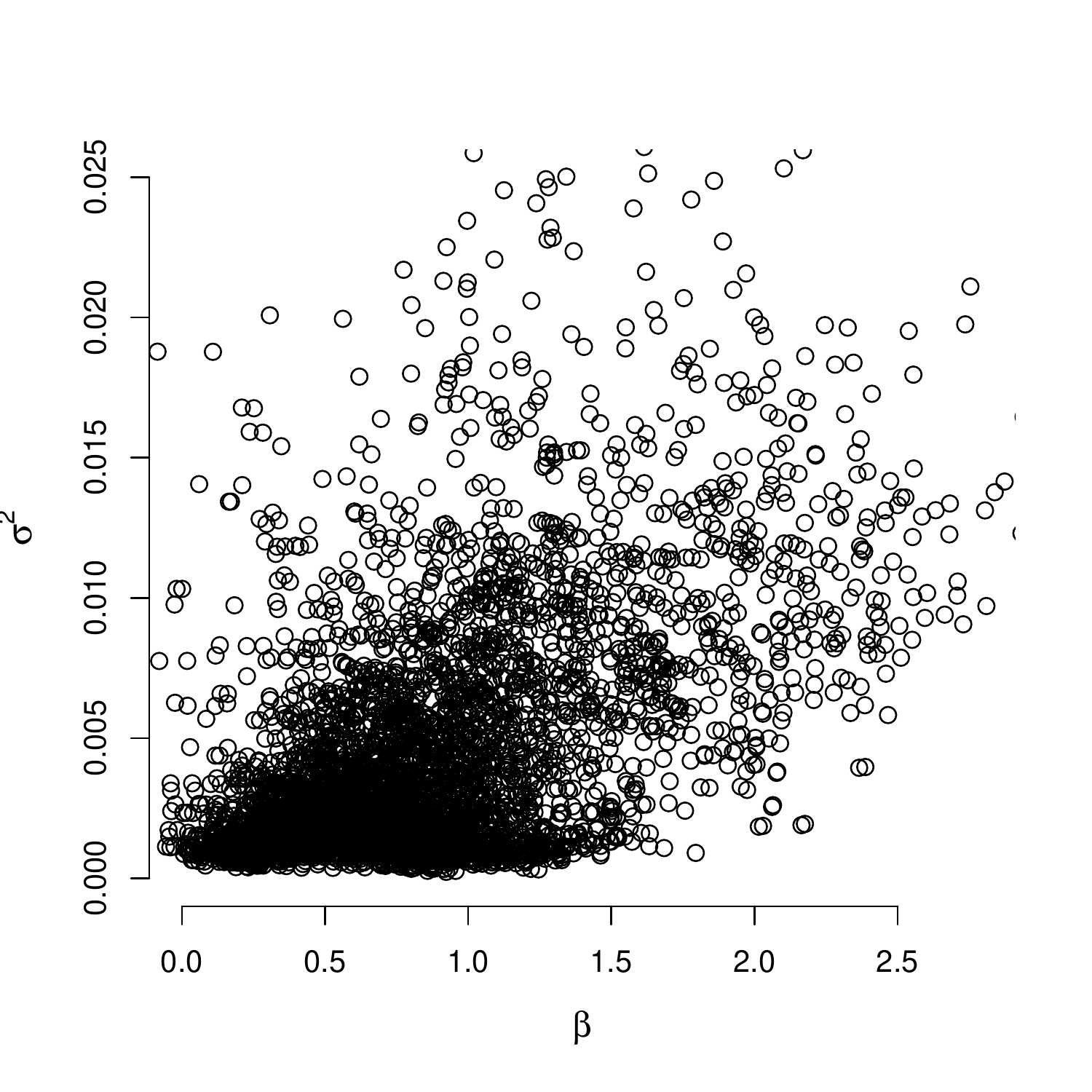} \includegraphics[width=0.4\linewidth]{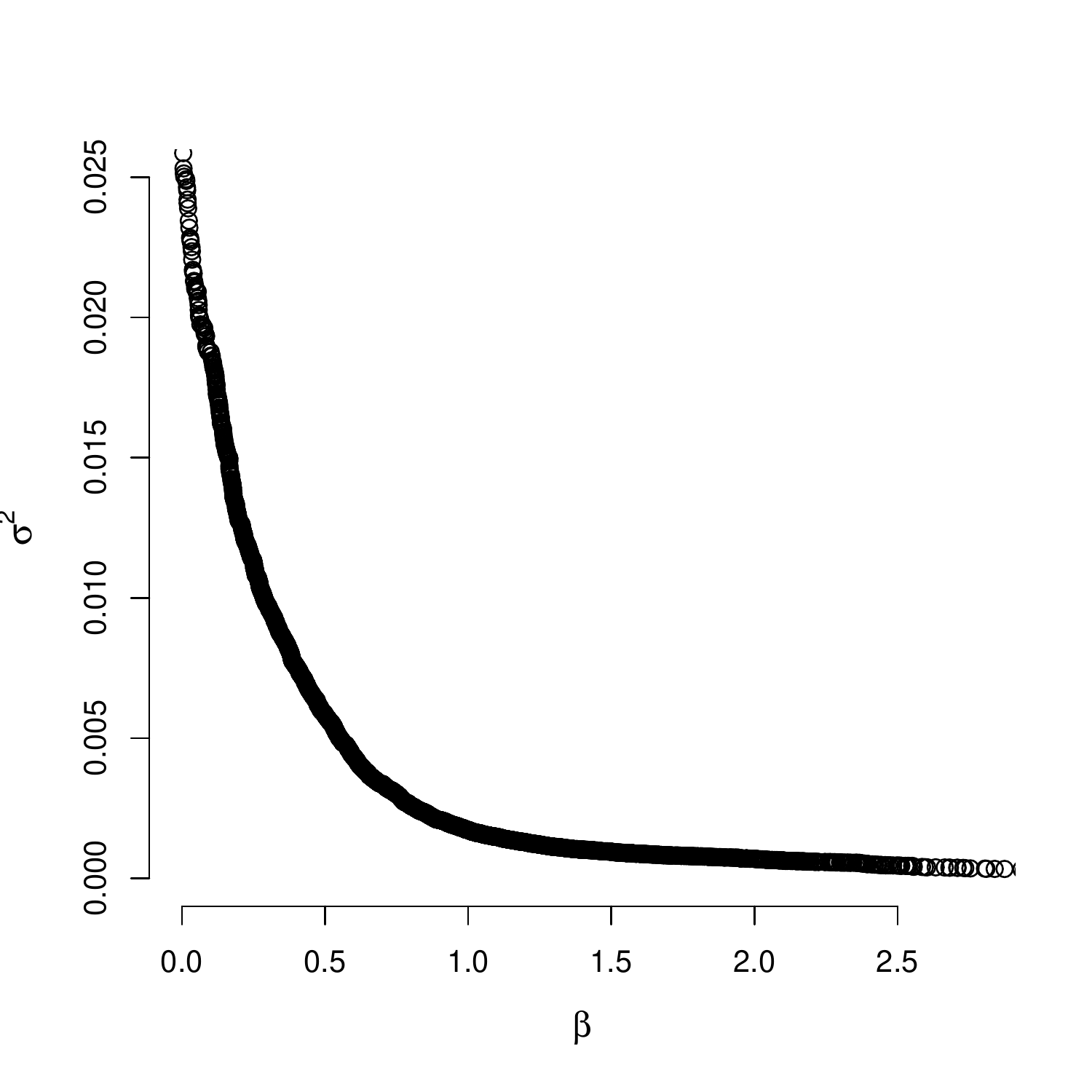}  \\ 
\caption{Indtrack5,6,7,8: Systematic ($\beta_i$) and nonsystematic ($\sigma_{\epsilon_i}$) Risk}
\end{figure}


\begin{table}[htb]
\extrarowheight = -1.3ex
\centering
\footnotesize
\begin{tabular}{lllrrrrrrr} 	\hline	
$N=4151$ &&&&	\multicolumn{5}{c}{Solution}	&		\\   
 \cline{1-3}  \cline{5-9}	&& $Model$	&&	$time$	&	$obj$	&	$\%desv$	&	K			&$\displaystyle	||w-w^*||_1$	&	\\
  \cline{1-3}  \cline{5-9}		
$K=5$  &&	$CCMVSF$		&&	3600.00	&	0.030696	&	&	5	&	&	\\								
&&	$CCMVSF_{LA}$			&&	20.01	&	0.030205	&	-1.60	\%	&	5	-	0	&	2.00	&	\\
&&	$EWCCMVSF$	&&	0.21	&	0.030203	&	-1.61	\%	&	5	- 2		&	1.20	&		\\		
&&	Alg.	(\ref{heuristic}) + Alg. (\ref{improving}) 		&&	0.03	&	0.030203	&	-1.61	\%	&	5	- 2		&	1.20	&		\\	\hline
$K=10$  &&	$CCMVSF$		&&	3600.00	&	0.025344	&	&	10	&	&	\\								
&&	$CCMVSF_{LA}$			&&	5.48	&	0.025184	&	-0.63	\%	&	10	- 2		&	1.61	&	\\
&&	$EWCCMVSF$	&&	0.43	&	0.025180	&	-0.65	\%	&	10	-	4	&	1.21	&		\\		
&&	Alg.	(\ref{heuristic}) + Alg. (\ref{improving}) 		&&	0.03	&	0.025196	&	-0.58	\%	&	10	- 4		&	1.21	&		\\	\hline
$K=20$  &&	$CCMVSF$		&&	3600.00	&	0.022003	&	&	20	&	&	\\								
&&	$CCMVSF_{LA}$			&&	149.73	&	0.020957	&	-4.75	\%	&	20	-	0	&	2.00	&	\\
&&	$EWCCMVSF$	&&	0.30	&	0.020957	&	-4.76	\%	&	20	- 0		&	2.00	&		\\		
&&	Alg.	(\ref{heuristic}) + Alg. (\ref{improving}) 		&&	0.03	&	0.020960	&	-4.74	\%	&	20 - 0		&	2.00	&		\\	
 \hline
 $K=30$  &&	$CCMVSF$		&&	3600.00	&	0.018621	&	&	30	&	&	\\								
&&	$CCMVSF_{LA}$			&&	286.09	&	0.018624	&	0.02	\%	&	30	-	10	&	1.33	&	\\
&&	$EWCCMVSF$	&&	0.25	&	0.018624	&	0.01	\%	&	30	- 11		&	1.22	&		\\		
&&	Alg.	(\ref{heuristic}) + Alg. (\ref{improving}) 		&&	0.03	&	0.018624	&	0.01	\%	&	30 - 11		&	1.22	&		\\	
 \hline
 $K=40$  &&	$CCMVSF$		&&	3600.00	&	0.016955	&	&	40	&	&	\\								
&&	$CCMVSF_{LA}$			&&106.10		&	0.016962	&		0.04\%	&	40	-	32	& 0.41		&	\\
&&	$EWCCMVSF$	&&	0.23	&	0.016954	&		-0.01\%	&	40	- 34		&	0.30	&		\\		
&&	Alg.	(\ref{heuristic}) + Alg. (\ref{improving}) 		&&	0.03	&	0.016954	&		-0.01\%	&	40 - 34		&	0.30	&		\\	
 \hline
 $K=50$  &&	$CCMVSF$		&&	3600.00	&	0.015727	&	&	50	&	&	\\								
&&	$CCMVSF_{LA}$			&&	26.18	&	0.015746	&		0.12\%	&	50	-	37	&	0.57	&	\\
&&	$EWCCMVSF$	&&	0.26	&		0.015745&		0.11\%	&	50	- 35		&	0.63	&		\\		
&&	Alg.	(\ref{heuristic}) + Alg. (\ref{improving}) 		&&	0.03	&	0.015745	&		0.12\%	&	50 -  40 &		0.43 &		\\	\hline
 \hline
 \multicolumn{9}{l}{Time limit 3600 sec.}
\end{tabular}
 \caption{indtrack5,6,7,8.txt \quad  \, N=4151}  	
\label{indtrack678}
 \end{table}

\end{document}